\documentclass[useAMS,usenatbib,usegraphicx]{mn2e}
\usepackage{aas_macros,times}

\title[ROLES z$\sim$1 SSFR--mass]
  {A spectroscopic measurement of galaxy formation timescales with ROLES
  }
\author[D. G. Gilbank et al.]
  {David G.~Gilbank$^1$\thanks{Email: dgilbank@astro.uwaterloo.ca}, 
Richard G.~Bower$^2$,  Karl~Glazebrook$^3$, 
Michael L.~Balogh$^1$, 
  \newauthor
 I.K.~Baldry$^4$,  G.T.~Davies$^2$,  
  G.K.T.~Hau$^{3}$, I.H.~Li$^3$, P. ~McCarthy$^5$, M.~Sawicki$^6$
\newauthor\\
  $^1$Department of Physics and Astronomy, University of Waterloo, Waterloo, Ontario, Canada N2L 3G1\\
    $^2$Institute for Computational Cosmology, Department of Physics, University of Durham, South Road, Durham, DH1 3LE, UK\\
  $^3$Centre for Astrophysics and Supercomputing, Swinburne University of Technology, P.O. Box 218, Hawthorn, VIC 3122, Australia\\
  $^4$Astrophysics Research Institute, Liverpool John Moores University, Twelve Quays House, Egerton Wharf, Birkenhead CH41 1LD, UK\\
  $^5$Carnegie Observatories, 813 Santa Barbara Street, Pasadena, California, 91101 USA\\
 $^6$Department of Astronomy and Physics, Saint Mary's University, Halifax, NS B3H 3C3, Canada\\
}

\pagerange{\pageref{firstpage}--\pageref{lastpage}} \pubyear{2006}

\def\LaTeX{L\kern-.36em\raise.3ex\hbox{a}\kern-.15em
    T\kern-.1667em\lower.7ex\hbox{E}\kern-.125emX}

\def\oii{[{\sc O\,II}]}

\def\g09{G10}

\def\ms{$\log(M_*/M_\odot)$}
\def\ha{{\rm H$\alpha$}}

\def\msunyr{M$_{\odot}$ yr$^{-1}$}

\def\lsim{\mathrel{\hbox{\rlap{\hbox{\lower4pt\hbox{$\sim$}}}\hbox{$<$}}}}
\def\gsim{\mathrel{\hbox{\rlap{\hbox{\lower4pt\hbox{$\sim$}}}\hbox{$>$}}}}

\begin{document}

\label{firstpage}

\maketitle

\begin{abstract}
We present measurements of the specific star-formation rate (SSFR)--stellar mass relation for star-forming galaxies. Our deep spectroscopic samples are based on the Redshift One LDSS3 Emission line Survey, ROLES, and European Southern Observatory, ESO, public spectroscopy at z$=$1, and on the Sloan Digital Sky Survey (SDSS) at z$=$0.1. These datasets cover an equally deep mass range of 8.5$\lsim$ \ms $\lsim$11 at both epochs. We find that the SSFR--mass relation evolves in a way which is remarkably independent of stellar mass, as we previously found for the star-formation rate density (SFRD)--mass relation. However, we see a subtle upturn in SSFR--mass for the lowest mass galaxies (which may at least partly be driven by mass-incompleteness in the K-selected sample). This upturn is suggestive of greater evolution for lower mass galaxies, which may be explained by less massive galaxies forming their stars later and on longer timescales than higher mass galaxies, as implied by the `cosmic downsizing' scenario. Parameterising  the e-folding timescale and formation redshift as simple functions of baryonic mass gives best fit parameterisations of $\tau(M_b) \propto M_b^{-1.01}$ and $1+z_f(M_b) \propto M_b^{0.30}$. This subtle upturn is also seen in the SFR density (SFRD) as a function of stellar mass.  At higher masses, such as those probed by previous surveys, the evolution in SSFR--mass is almost independent of stellar mass. At higher masses (\ms$>$10) the shapes of the cumulative cosmic SFRDs are very similar at both z$=$0.1 and z$=$1.0, both showing 70\% of the total SFRD above a mass of \ms$>$10. Mass functions are constructed for star-forming galaxies and found to evolve by only $<$35\% between z$=$1 and z$=$0.1 over the whole mass range. The evolution is such that the mass function decreases with increasing cosmic time, confirming that galaxies are leaving the star-forming sequence/blue cloud. The observational results are extended to z$\sim$2 by adding two recent Lyman break galaxy samples, and data at these three epochs (z$=$0.1, 1, 2) are compared with the {\sc galform} semi-analytic model of galaxy formation. {\sc galform} predicts an overall SFR density (SFRD) as a function of stellar mass in reasonable agreement with the observations. The star formation timescales inferred from 1/SSFR also give reasonable overall agreement, with the agreement becoming worse at the lowest and highest masses. The models do not reproduce the SSFR upturn seen in our data at low masses, where the effects of extinction and AGN feedback should be minimal and the comparison should be most robust.

\noindent
\end{abstract}

\begin{keywords}
galaxies: dwarf --
galaxies: evolution --
galaxies: general
\end{keywords}

\section{Introduction}
\label{sec:introduction}
A useful metric for quantifying activity in a galaxy is the specific SFR (SSFR, SFR per unit stellar mass, SFR/M$_\star$). This represents an efficiency of star-formation, since it measures the observed star-formation rate relative to that which it must have had in the past in order to build up the observed amount of stellar mass (e.g., \citealt{Kennicutt:1994ud,Brinchmann:2004ct}). In the local Universe, a correlation between the SSFR and stellar mass is observed \citep{Brinchmann:2004ct}, such that lower mass galaxies exhibit higher SSFRs than their higher mass counterparts. A similar trend is also seen at higher redshifts \citep{Brinchmann:2000th,Feulner:2005la,Elbaz:2007rt,Noeske:2007tw,Zheng:2007vl,Damen:2009gd}, with the overall normalisation of this relation shifting to higher SSFRs at earlier cosmic times. One result of this evolving relation is that the number of galaxies exceeding some SSFR threshold shifts from higher to lower mass galaxies with increasing cosmic time, generally referred to as `cosmic downsizing' \citep{Cowie:1996xw}, or `downsizing in (S)SFR'\footnote{To distinguish it from other observations which point to this scenario from different avenues (such as the observation that locally-observed massive galaxies exhibit older stellar populations than their less massive counterparts `archaeological downsizing' \citep{Kauffmann:2003qm,Thomas:2005ca}). See \citet{Fontanot:2009uq} for a comprehensive summary.}.

The evolution of this relation between SSFR and mass represents an important test of galaxy formation models, since both SFR and stellar mass depend critically on the prescriptions used for star-formation and mechanisms (``feedback") which act to suppress it. Thus, observations which measure how SSFR and stellar mass are related and evolve with cosmic time are required to constrain such models. Several observational limitations to building such a study exist. Possibly the most serious of these is the use of different SFR indicators which have different sensitivities to unobscured (e.g., UV, \oii, \ha) and obscured star-formation (e.g., far infra-red emission, FIR); additional dependencies, such as metallicity (e.g., \oii); and contamination from non-starforming sources. The need to span a wide redshift range often means that different indicators must be used at different redshifts within a given survey, necessitating switching between different indicators, in which case systematic differences between the indicators might masquerade as evolution.  

Previous studies of SSFR--mass and its evolution have used either mixtures of different SFR indicators (e.g., \citealt{Juneau:2005ft,Noeske:2007tw}), photometric redshifts which require SFR, redshift and dust extinction, etc. to be estimated, in an often degenerate way, from the same set of data (e.g., \citealt{Feulner:2005la,Damen:2009it,Zheng:2007vl}). Many studies also use stacking techniques, in which case contamination from a subset of objects (such as AGN) is often difficult to detect, and suffers from the limitation that only the average property of the stacked sample may be estimated (e.g., \citealt{Zheng:2007vl,Dunne:2009bg,Pannella:2009uj}). Those which use spectroscopy to obtain redshifts (and/or spectroscopic SFRs) and consistent indicators at all redshifts still only probe the most massive galaxies at higher redshifts \citep{Elbaz:2007rt,Cowie:2008ob,Maier:2009sf}
 
In \citet[][hereafter Paper 2]{Gilbank:2010hc} we presented a new survey (the Redshift One LDSS3 Emission lines Survey, ROLES) designed to target with mulit-object spectroscopy (MOS) galaxies an order-of-magnitude less massive than previously studied at z$\sim$1. We advocated the use of \oii~as a SFR indicator, empirically-corrected as a function of galaxy stellar mass in order to correct for extinction and other systematic effects with using \oii. This is useful as \oii~is more easily accessible to optical MOS instruments out to moderate redshifts z$\sim1.2$ and thus leads to relatively efficient surveys. We showed by comparison with Balmer decrement-corrected \ha~SFR that this mass-dependent correction is reliable locally \citep{Gilbank:2009nx} and furthermore gives good agreement with the extinction estimated from IR measures. 

In this paper we construct a sample using a single SFR indicator from $0\lsim z \lsim 1$\footnote{The results we present actually use \ha~at z$=$0.1 to combat incompleteness for \oii~at high stellar masses in SDSS. For the low mass galaxies which are the primary aim of ROLES, our results are unchanged if we use \oii~directly instead of \ha~at z$=$0.1 (c.f. \citealt{Gilbank:2009nx}).}, spanning the widest possible mass range studied with spectroscopy (for redshifts and SFRs) over this full range. By combining our low mass spectroscopic survey at z$=$1 (Paper 2) with higher mass data from the literature, and  our local comparison sample built from SDSS \citep{Gilbank:2009nx}, we construct an unprecedented sample in this parameter space of mass--redshift--SFR. In \S2 we present the samples used; \S3 presents our results for the SSFR--mass relation at z$=$0.1 and z$=$1, and for the mass functions of star-forming galaxies. \S4 uses a toy model to describe the behaviour seen in the SSFR--mass relation, and compares several of our observational results with predictions from the {\sc galform} semi-analytic model of galaxy formation. A discussion of possible systematic errors is made in \S5 and the conclusions are presented in \S6. Throughout we assume a $\Lambda$CDM cosmology with $\Omega_m=0.3$, $\Lambda=0.7$, and $H_0=70$\,km s$^{-1}$ Mpc$^{-1}$. All magnitudes use the AB system. All stellar masses and SFRs are calculated assuming (or transforming to - see Appendix \ref{sec:imf}) a \citet[][hereafter BG03]{Baldry:2003ue} initial mass function (IMF).

\section{Sample}
\label{sec:data}

\subsection{z$=$1 data}
In order to minimise uncertainties caused by comparing different SFR indicators, a z$\sim$1 sample is constructed from surveys using emission line indicators for objects with spectroscopic redshifts. $0.88<z\le1.15$ data are taken from the sample described in Paper 2. Of the two ROLES fields studied in paper 2 (the Faint Infra-Red Extragalactic Survey, FIRES, and the {\it Chandra} Deep Field South, CDFS) only the CDFS is considered here. This is due to the wealth of additional data available in the CDFS (including additional public spectroscopy covering higher mass galaxies than probed by ROLES alone). In addition, the CDFS results dominate the ROLES statistics (199 low mass galaxies vs 86 for FIRES) due to the larger volume probed by the CDFS data. Briefly these data comprise multicolour photometry from FIREWORKS \citep{Wuyts:2008lq}, and spectroscopic redshifts and \oii~SFRs from ROLES (Paper 2) for low mass (8.5$\lsim$ \ms $\lsim$9.5) galaxies. ROLES is a highly-complete, low-mass survey with a sampling completeness $>$80\% comprising 199 galaxies in the mass range 8.5$<$ \ms $\le$9.3 (Paper 2). The SFR is estimated from the \citet{Kennicutt:1998pa} relation (which assumes 1 mag of extinction at \ha~and an \oii/\ha~ratio of 0.5) converted to our BG03 IMF. This is then corrected as a function of mass using the empirical mass-dependent correction of \citet{Gilbank:2009nx}. Stellar masses are determined by SED-fitting of the photometry, at the spectroscopically-determined redshift, using a grid of PEGASE.2 models and a BG03 IMF as described in \citet{Glazebrook:2004zr}.

The high mass end in CDFS is supplemented with ESO public spectroscopy \citep[][hereafter referred to as the `FORS2' data]{Vanzella:2008vq} for which we have measured \oii~fluxes from their 1D spectra and computed masses using exactly the same FIREWORKS photometry as used for ROLES (Paper 2). The FORS2 sample is less complete than ROLES, comprising 73 galaxies in the mass range 9.7 $<$ \ms $\le$11.2 and an average sampling completeness of $\sim$30-40\% (Paper 2) with a typical redshift success rate of 72\% \citep{Vanzella:2008vq}. 

\begin{figure}
	{\centering
	\includegraphics[width=85mm,angle=0]{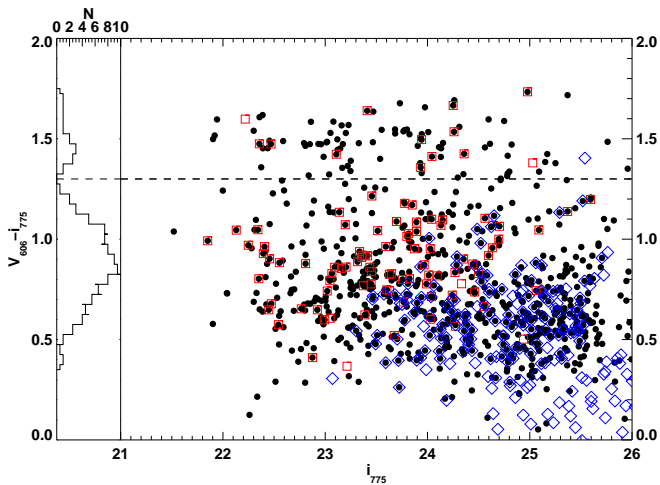}
	\caption{$(V_{606}-i_{775})$ -- $i_{775}$ colour-magnitude diagram for galaxies with photometric redshifts at z$\sim$1 (filled circles). Red open squares indicate galaxies with spectroscopic redshifts in the ROLES' redshift range from the FORS2 sample, and open blue diamonds indicate spectroscipically-confirmed z$\sim$1 galaxies from ROLES spectroscopy. The left panel shows a histogram of the colours, clearly indicating the bimodal nature of the galaxy colour distribution at z$\sim$1. The dashed horizontal line at $V_{606}-i_{775}=1.3$ indicates the division between star-forming (blue cloud) and passive (red-sequence) galaxies. }
	 \label{fig:cmd}
}
\end{figure}

We explore the effect of selecting star-forming galaxies by their blue colours. This is often done in spectroscopic samples to reject possible contamination from emission line objects where the emission is not from star-formation, such as LINERS \citep{Yan:2006wd}, and might be done in purely photometric surveys in order to reject non star-forming galaxies (see also \S\ref{sec:sampsel}). For these CDFS samples, the blue cloud is isolated by examining a colour bracketing the 4000\AA~break for galaxies with photometric redshifts (photo-z's) from FIREWORKS within the ROLES redshift range of $0.88<z\le1.15$. Fig.~\ref{fig:cmd} shows the $(V_{606}-i_{775})$ -- $i_{775}$ colour-magnitude diagram (CMD) for galaxies in the CDFS with mean photometric redshifts in the ROLES' redshift window (black points). The left panel shows a histogram of the $(V_{606}-i_{775})$ colour which is clearly bimodal. The dashed horizontal line at $(V_{606}-i_{775})=1.3$ indicates a dividing line between red and blue galaxies. The exact position of this line is not critical and any reasonable cut around this value leads to comparable results. Open red squares indicate galaxies with spectroscopic redshifts in our redshift window from the FORS2 subsample. The majority of these objects inhabit the blue cloud, but there is also a significant subsample residing on the red-sequence. This latter category may comprise objects with very low levels of star-formation (since we have pre-selected objects with detectable \oii~emission), not detectable from broad-band colours, or possibly emission unrelated to star-formation such as LINERs \citep{Yan:2006wd}. Rejecting red-sequence galaxies in this way reduces the sample size somewhat, to 60 rather than 73. For consistency, we recompute the completenesses in the same way as Paper 2, this time considering the blue sample separately and find that this makes negligible difference (except in the brightest $K$ bin which was previously 70\% complete), the overall completeness remaining at $\sim$30-40\% independent of $K$-band magnitude. 

Blue diamonds indicate z$\sim$1 ROLES galaxies.\footnote{It should be noted that some of the spectroscopic redshift slice-members (open symbols) do not correspond to galaxies with a best-fit photometric redshift in this slice (filled circles). The photo-z selected galaxies here only consider the value of the most-likely redshift, as we just require a representative z$\sim$1 CMD to select red vs blue galaxies. In the analysis to assign spectroscopic redshifts, we consider the full photo-z PDFs (paper 2) and so galaxies with peak photo-z's outside this window may still have a significant probability of belonging to this slice.} As can be seen, all but one object clearly belongs to the blue sequence (as might be expected given our selection criteria), so the ROLES sample may effectively be considered as a blue cloud sample. Indeed, the lack of red objects in ROLES is strongly suggestive that these low mass galaxies possess little dust.

\subsubsection{Mass completeness of z$\sim$1 data}
\label{sec:masscompl} 

\begin{figure}
	{\centering
		\includegraphics[width=85mm,angle=0]{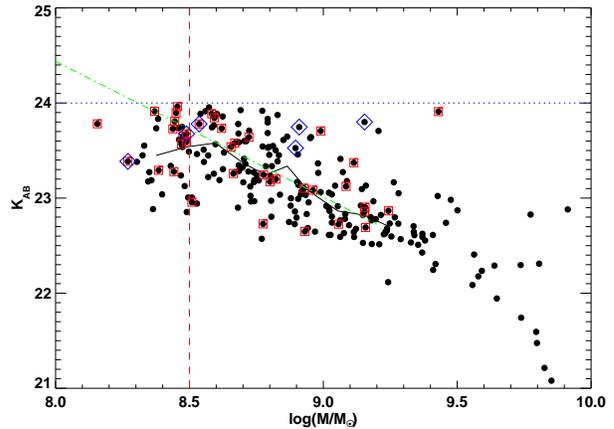}
	\caption{Mass vs observed $K$-band magnitude for the z$\sim$1 sample. Black points show all galaxies in the ROLES$+$FORS2 samples. In this work, only galaxies with masses down to \ms$=$8.5 (vertical dashed red line) are considered and brighter than $K$=24.0 (horizontal dotted blue line). Galaxies with \oii-SFR$\le$0.3\msunyr~ are indicated by red diamonds. Galaxies with SED-fit SFR$\le$0.3\msunyr~are indicated by blue squares. Solid black line shows running median of 15 galaxies in \ms--$K$ and dotted green line shows an extrapolation of this relation where the data are clearly complete. See text for discussion.  }
	 \label{fig:masscomp}
}
\end{figure}

Estimating the completeness in mass for this low mass galaxy sample is difficult, since the spectroscopy is pushing the limit of the optical and NIR photometry. Our results probe down to masses of \ms$=$8.5 at our magnitude limit of $K$=24.0. So, it is important to check that $K$=24.0 is bright enough to sample the z$\sim$1 star-forming galaxies of interest. In Paper 2, we used photometric redshifts to show that galaxies potentially within our redshift range with SFRs below our \oii~selection threshold likely do not significantly contribute to the global SFRD at z$\sim$1. Pushing these photo-z's to fainter fluxes to attempt to locate galaxies below our $K$=24.0 limit with stellar masses potentially within our mass window is complicated by the rapidly increasing photometric errors below this limit. However, instead of resorting to the photo-z's we can use the spectroscopic results to search for potential selection biases within our sample.  Fig.~\ref{fig:masscomp} shows the observed $K$-band magnitude as a function of fitted stellar mass for all objects in the initial ROLES$+$FORS2 sample (i.e., prior to applying magnitude and mass cuts). The general trend of $K$ versus stellar mass (and its scatter) can be seen. At masses above \ms$\sim$9, the distribution is well-separated from the K$=$24 magnitude limit (horizontal dotted line). Thus, a linear extrapolation of the (median-smoothed) relation around this mass towards the \ms$=$8.5 limit gives some indication of how the average relation should look in the absence of any possible selection bias. The average relation (measured by a running median of 15 objects) agrees reasonably well with the linear extrapolation from brighter magnitudes, indicating that the bulk of objects at the \ms$=$8.5 mass limit are likely above the $K$=24 magnitude limit. Assuming the scatter in magnitude at a given mass remains constant, a visual estimate from this plot suggest that mass incompleteness at the magnitude limit will be minimal. Attempting to quantify this, fading galaxies at \ms$\sim$9.5 to \ms$=$8.5  assuming the same mass-to-light ratio (M/L) as the higher mass galaxies, a maximum of 25\% of these objects would fall below the $K$-band magnitude cut-off. Thus we may estimate that we are at least 75\% complete at the very lowest mass limit of ROLES. 

In order to check that our sample is not biased towards more actively star-forming galaxies near the magnitude limit of the survey, the lowest \oii-SFR galaxies ($\le$0.3\msunyr) are indicated in Fig.~\ref{fig:masscomp}. It can be seen that these objects are distributed over a wide range in mass, and thus we do not see a `pile up' of more-actively star-forming galaxies at the mass/magnitude limit of the survey. This tallies with the difference in observed $K$-band magnitude at a given stellar mass from the extensive grid of PEGASE models used in the stellar mass fitting. At a given SFR for a model \ms$=$8.5 galaxy (the mass limit of ROLES), for a wide range of metallicities, extinctions, and star-formation histories, the intrinsic scatter in $K$ is $\approx$ 0.5 mag or a factor of about 60\% in M/L. The difference between a SFR of 0.3\msunyr~and 0.1\msunyr~(the former being the approximate ROLES limit and the latter thus being a factor of three below the nominal limit) corresponds to a systematic offset of 0.4 mag in observed $K$. Thus, while it is slightly easier to observe more actively star-forming galaxies in our sample, our empirical test shown in Fig.~\ref{fig:masscomp} suggests that we are not biased in this way.

\subsection{Local (z$=$0.1) data}
Local (z$\sim$0.1) data are taken from the SDSS Stripe 82 sample of \citet{Gilbank:2009nx}. This sample was cut in redshift to $0.032 < z < 0.20$ to ensure the inclusion of the \oii~line at the low redshift end, and to avoid the effects of incompleteness/evolution at the high redshift end. For the present work we switch from using the \oii-derived estimates of SFR to using the Balmer decrement-corrected \ha~measurements directly. \citet{Gilbank:2009nx} showed that the \oii-SFRs could be empirically corrected to agree statistically with the \ha-derived SFRs, but that the depth of the SDSS spectroscopy led to incompleteness for high mass galaxies when \oii~was used as the SFR indicator, due to the reduced sensitivity of \oii~to SFR with increasing M$_\star$. Since this work primarily requires high completeness, we choose to adopt \ha~as our indicator at z$\sim$0, safe in the knowledge that it agrees, on average, with mass dependent empirically-corrected \oii-SFR. To select galaxies belonging to the blue cloud, we adopt the rest-frame $(u-g)$ colour cut of \citet{Prescott:2009vz} to separate red and blue galaxies, as used in \citet{Gilbank:2009nx}.

\subsection{Specific star-formation rates}
As mentioned above, the SFR per unit stellar mass (specific SFR, SSFR) is a useful quantity for gauging the efficiency of star-formation. The inverse of SSFR defines a timescale for star-formation, i.e., 
\begin{equation}
SSFR^{-1} \propto t_{SFR} \propto M_\star/\dot{M}_\star 
\label{eqn:invssfr}
\end{equation}
(where the SFR is the time derivative, $\dot{M_*}$, of the stellar mass, $M_*$), 
which is simply the time required for the galaxy to form its stellar mass, assuming its SFR remained constant. In the following, we will be considering the SSFR, and closely related quantities, of ensembles of galaxies. It is important to distinguish between the way the different quantities are calculated, depending on whether one is interested in the properties of the average galaxy, or the cosmic average of all galaxies. These distinctions are nicely described in \citet{Brinchmann:2004ct} where they use $r_{SFR}^g$ to denote the SSFR ($r_{SFR}$) of the typical galaxy, and $r_{SFR}^V$ to denote the volume-averaged equivalent. These can be written 
\begin{equation}
r_{SFR}^g = \displaystyle \left< \frac{ {\mathrm SFR} }{M_\star} \right> = \displaystyle \left< \frac{ \Psi }{M_\star} \right>
\label{eqn:r_g}
\end{equation}
and
\begin{equation}
r_{SFR}^{V} = \frac{\rho_{SFR}}{\rho_{M_\star}}. 
\label{eqn:r_v}
\end{equation}
Volume-averaged quantities (Eqn.~\ref{eqn:r_v}) are calculated following the method described in Paper 2. Briefly, this involves using the Vmax technique and weighting by sampling completeness estimated from detailed photometric redshift probability distribution functions.

\section{Results}
\subsection{The SSFR--mass relation}
\label{sec:ssfrmass}
\begin{figure*}
	{\centering
		\includegraphics[width=85mm,angle=0]{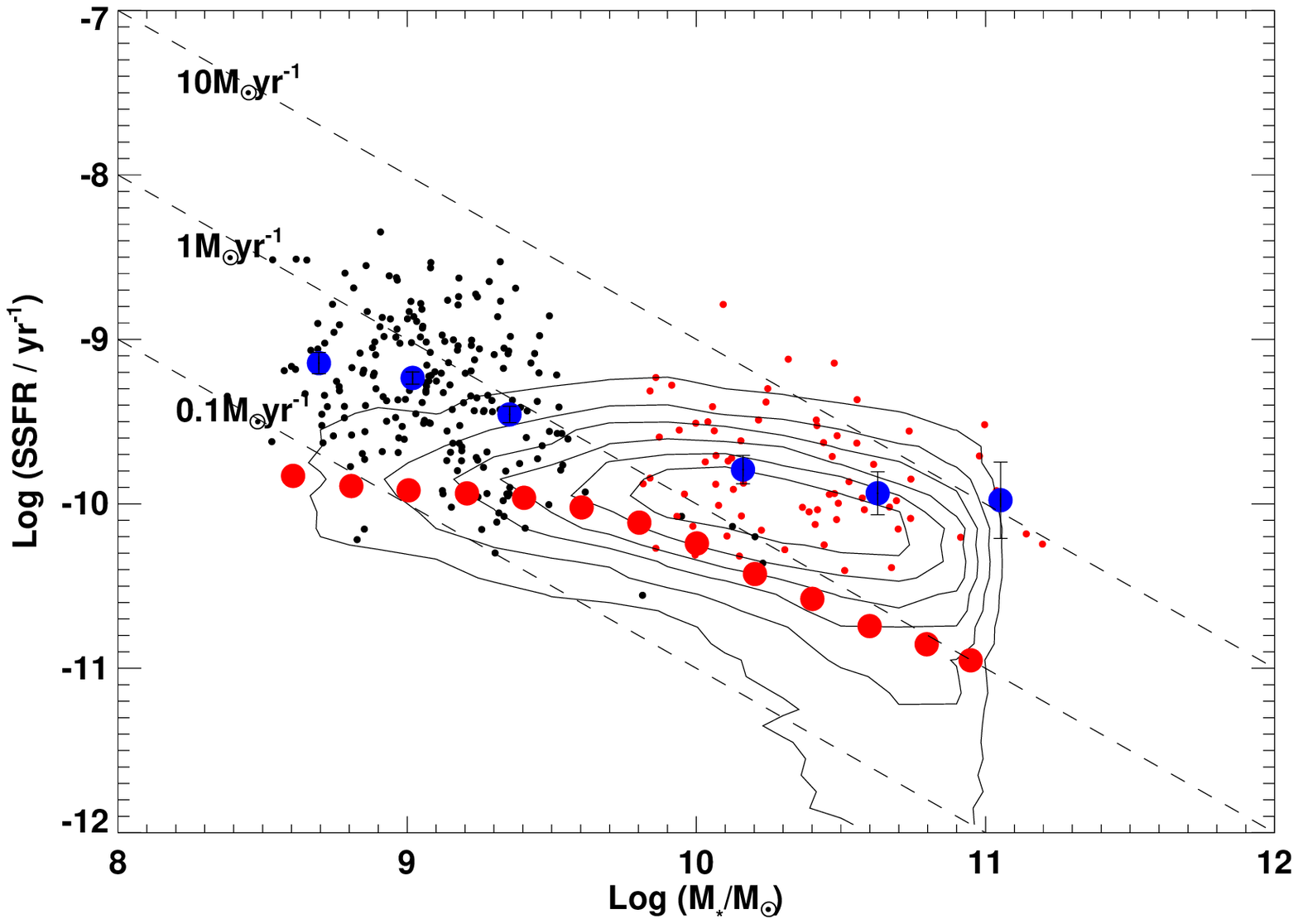}
		\includegraphics[width=85mm,angle=0]{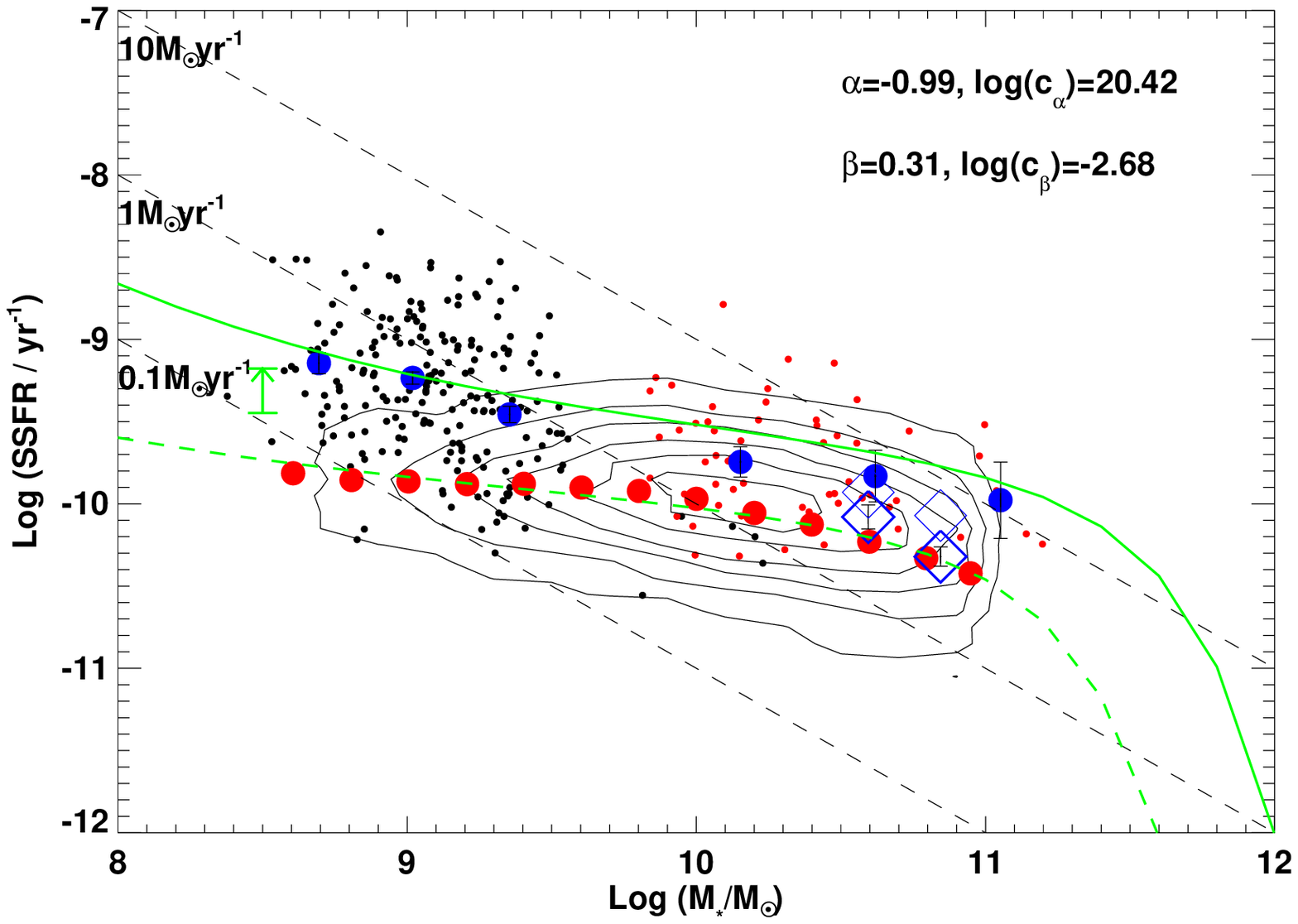}
	\caption{$\log$(SSFR) versus $\log$(stellar mass) for galaxies in our samples. The left panel shows data for galaxies of all colours (with significant emission lines) in the spectroscopc samples; the right panel shows galaxies after application of the blue colour cut. Small, filled points show data at $0.88<z\le1.15$: black points are from ROLES, the higher mass data (smaller red filled points) are taken from ESO (FORS2) public spectroscopy. Contours denote SDSS data at z$\sim$0.1. Larger filled circles show mean SSFRs in bins of stellar mass for the data in the two redshift ranges: blue symbols at z$\sim$1 and red symbols at z$\sim$0.1. Dashed lines denote the three different SFRs, as annotated.  In the right panel, the dotted black lines show best fit power laws to the z$=$1 data, as described in the text. Green lines show best-fit tau models, discussed in \S\ref{sec:tau}, with parameters as indicated on plot. The green arrow indicates the possible effect of incompleteness, showing the systematic shift (head of arrow) caused by removing the 25\% lowest SSFR points from the \ms$\sim$9.2 bin (tail of arrow). See text for details.}
	 \label{fig:ssfrmass}
}
\end{figure*}

SFRs and stellar masses from the z$\sim$0.1 and z$\sim$1 samples are combined to study the evolution of SSFR--mass in Fig.~\ref{fig:ssfrmass}. The left panel shows this dataset without any colour pre-selection, whereas the right panel shows the same data after first removing the red-sequence galaxies, as described above. The SDSS (z$\sim$0.1) galaxies are shown as contours for clarity. Filled red circles show the mean in bins of stellar mass. The z$\sim$1 data are shown as smaller filled circles, colour coded by dataset: black points are ROLES' galaxies; smaller red filled circles show the data from the public FORS2 spectroscopy in CDFS. Although individual data points are shown for each galaxy, the mean is calculated weighting each galaxy by a completeness and Vmax weight. Larger black circles show the median in bins of stellar mass for the z$\sim$1 data.  For the FORS2 data, the mean relation is not significantly lowered by including all galaxies with \oii~emission, rather than just blue cloud galaxies with \oii~emission. However, the colour cut makes a much larger difference to the mean SSFR measured in the SDSS. The effect of a colour cut to select star-forming galaxies will be discussed in \S\ref{sec:discuss}. For now, we note that we prefer the sample removing red-sequence members (i.e., the right panel) and proceed to consider this.

The dotted line shows the best-fit power law, $\log (SSFR) = -0.42 \log (M_\star) -5.49$, which is consistent with all the points within their 1$\sigma$ errors, although the data are suggestive of a stepping of the slope toward lower masses. The green lines indicate a toy `tau' model which will be described in \S\ref{sec:tau}.

In Fig.~\ref{fig:ssfrmass}, we see that the $\log($SSFR$)$--$\log($mass$)$ relation evolves to lower normalisation at lower redshifts (as found by many other works). At the high mass end, the relation evolves almost in parallel, as found by, e.g., \citet{Zheng:2007vl}. However, when the low mass ROLES data are included, an upturn in the relation at lower masses is seen. 

In order to verify that this upturn is significant, a bootstrap technique is used to perform 1000 realisations of the z$\sim$1 data. The bootstrap error bars for each data point are comparable with, or smaller than, the 1$\sigma$ sampling errors shown in the plot. Furthermore, if the approximate mid-points of the lower and higher mass samples, given by simply taking the centre-most bins of the ROLES and FORS2 data (i.e., \ms$=$9.0 and \ms$=$10.6), then the lower mass sample is always significantly higher than than the higher mass bin in all but two of the 1000 simulations. If a weighted mean of all three bins in the low and high mass samples is considered, then this would strengthen the significance further. The main uncertainty between the lower and higher mass data is likely to be systematic in nature (discussed in \S\ref{sec:discuss}), and so we also consider the following bootstrap test. The null hypothesis that the data in the lowest mass bin is in fact the same as that of the highest mass ROLES bin, \ms$\sim$9.2, (i.e., that the z$=$1 SSFR--mass relation is in fact flat and that the observed upturn is artificial) is adopted. Data points from the \ms$\sim$9.2 bin are taken and bootstrap resampled, with the lowest 25\% of these points removed each time (making the conservative assumption that the $\approx$ 25\% incompleteness, estimated in \S\ref{sec:masscompl}, systematically occurs for the lowest SSFR values). This generates a new mean SSFR value in each realisation which is obviously systematically higher than that without the lower points removed. In 1000 realisations, only seven are as high as the mean SSFR observed in the data for the lowest mass ROLES point. A limit to this bias may be estimated from the 95 percentile of the bootstrapped distribution. This is indicated by the green arrow in Fig.~\ref{fig:ssfrmass} and is approximately consistent with the lower 1$\sigma$ error bar of the lowest mass ROLES point. This shows that 5\% of the time, the upturn seen in our data may be produced from a flat relation, with the lowest mass and lowest SFR galaxies dropping below the $K$-band limit of our sample. We emphasize again that this test is likely conservative, since it assumes that the distribution of \ms$\sim$8.5 galaxies is exactly the same as those of \ms$\sim$9.2 galaxies, and that the 25\% lowest SSFR galaxies are removed in each bootstrap realisation. The fact that we measure an upturn in this data is therefore intriguing but not cannot be completely dismissed as due to selection effects, and can only be addressed with even deeper K-band data in future mass-selected samples.

The reason for this upturn will be discussed in \S\ref{sec:tau}. The low mass data should be the least uncertain in terms of total SFR estimated from \oii, since these objects possess little extinction, as shown in \S\ref{sec:stack24}. Finally, it is worth noting that the combination of SDSS plus the ROLES$+$FORS2 data spans the largest range in mass and redshift probed by a uniformly-selected, spectroscopic sample.

\subsection{The mass function of star-forming galaxies}
\label{sec:mf}
\begin{figure*}
	{\centering
	\includegraphics[width=85mm,angle=0]{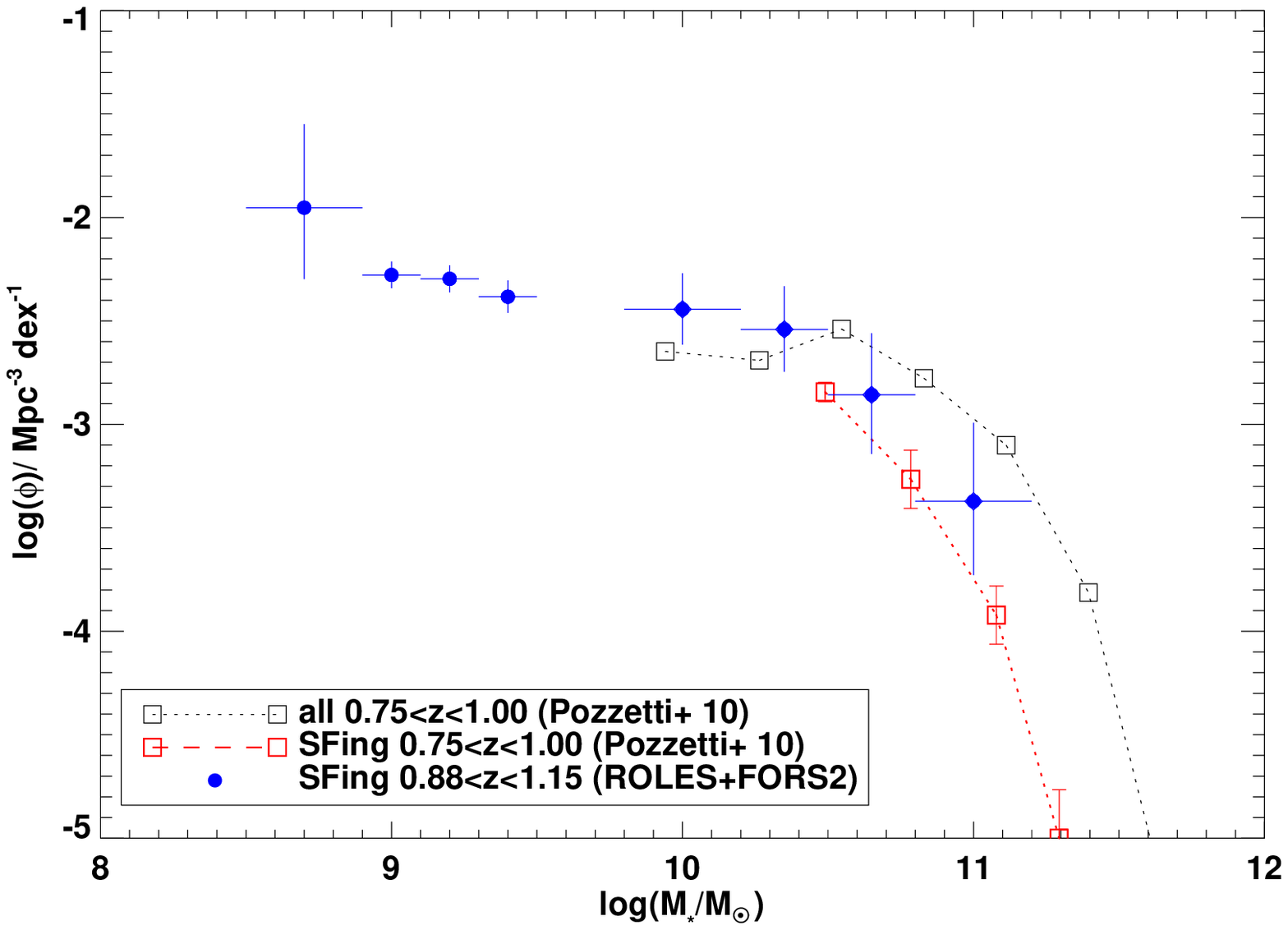}
		\includegraphics[width=85mm,angle=0]{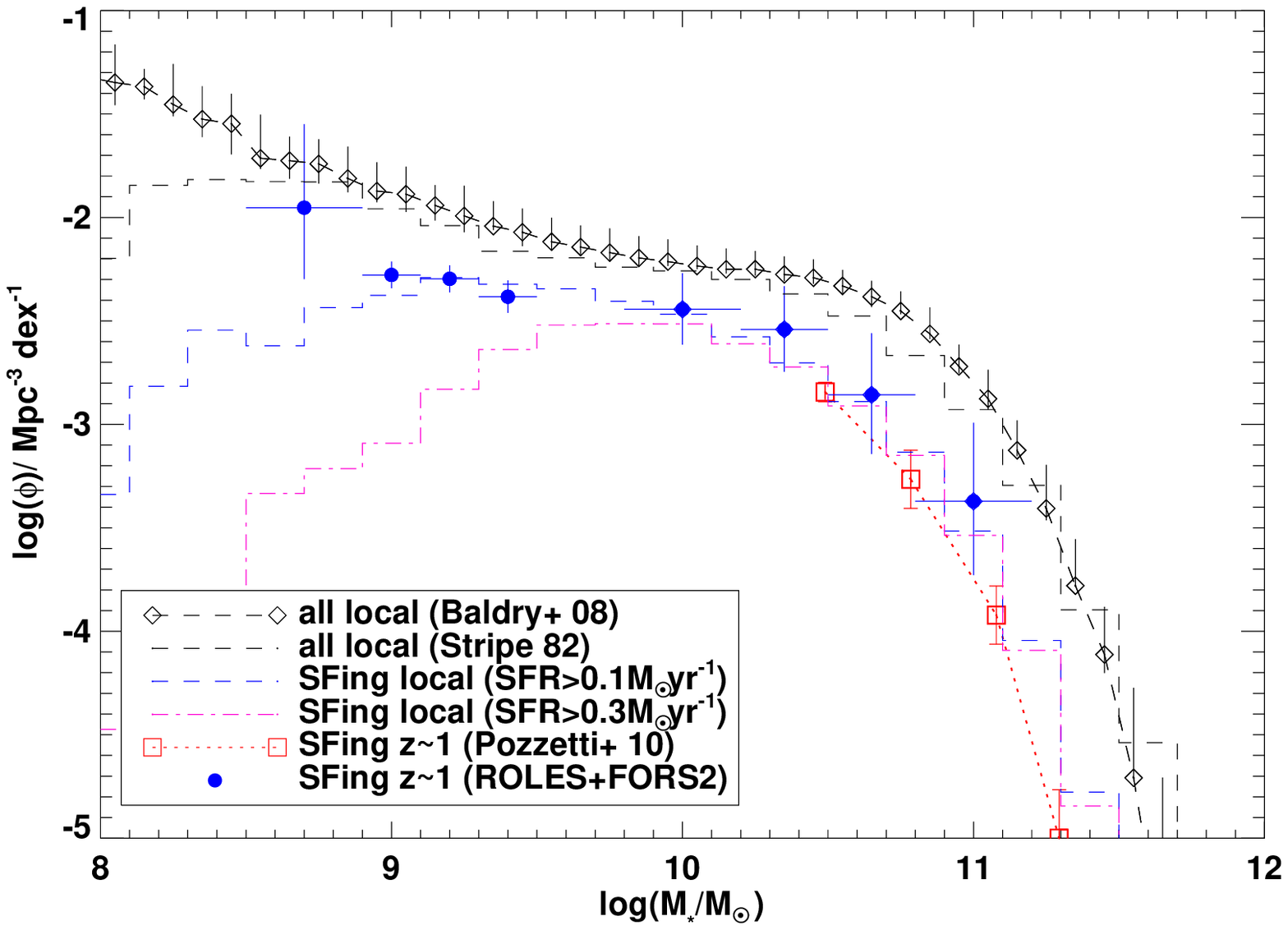}
	\caption{Left panel: mass functions at z$\sim$1. Black diamonds and squares show data for galaxies of all types from \citet{Drory:2005ts} and \citet{Pozzetti:2009qv} respectively. Red squares denote (blue) star-forming galaxies from \citet{Pozzetti:2009qv} and blue filled circles show (blue) star-forming galaxies from the CDFS from ROLES$+$FORS2 spectroscopy. Right panel: Comparison of the z$\sim$1 mass functions with local values calculated from the Stripe 82 sample of \citet{Gilbank:2009nx}. Black diamonds show the mass function for all galaxies, which  may be compared with that computed from the NYU-VAGC in \citep{Baldry:2008sj}. Blue and magenta lines indicate mass functions for star-forming galaxies (blue-selected) with SFRs$>$0.3 \msunyr~and $>$0.1\msunyr, respectively. Note that the depth of the ROLES data at z$=$1 reaches the same mass limit as in the local data.
	}
	 \label{fig:mf}
}
\end{figure*}

The preceding has examined the relation between SSFR and stellar mass. It is instructive to look at how this mass evolves with time. Since ROLES is an emission line-selected survey, it is not possible to measure the mass function for {\it all} galaxies at z$=$1 from this sample. However, it is possible to measure the mass function of {\it star-forming} galaxies using the same Vmax method used to calculate the SFRD (Paper 2, eqn.~18), replacing the SFR of each galaxy with unity. The number density of blue, star-forming galaxies, $\phi$, as a function of \ms~is shown in Fig.~\ref{fig:mf}. ROLES$+$FORS2 CDFS data are shown as filled blue circles with error bars. Open red squares show the mass function of blue galaxies (0.75$<$z$\le$1.00) from \citet{Pozzetti:2009qv} using zCOSMOS spectroscopic redshifts.  The left panel also shows  the mass function for galaxies of {\it all} colours (0.75$<$z$\le$1.00) from the  zCOSMOS spectroscopic data of \citet{Pozzetti:2009qv} (open squares). 

\citet{Pozzetti:2009qv} classifiy star-forming galaxies by their photometric type from SED-fitting at their spectroscopic redshift (which is not quite the same as selecting star-forming galaxies by their \oii~emission and blue colours, as we have done). The results shown in Fig.~\ref{fig:mf} uses their SED-fit late type galaxies (SED-LTG) class (from their fig.~11), which should most closely resemble our blue star-forming selection. Our results are consistent with theirs, within the errors. Our points are systematically below theirs, but this small offset is consistent with cosmic variance expected from the CDFS volume (Paper 2), or this could represent  differences in the selection criteria.

The right panel repeats a subsample of the z$=$1 results and shows local comparison data. Open diamonds with error bars show the local galaxy mass function for galaxies of all types from the NYU-VAGC subsample of SDSS from \citet{Baldry:2008sj}. The black dashed histogram shows the equivalent mass function from the Stripe 82 sample \citep{Gilbank:2009nx} computed for the present work. The good agreement between the two datasets taken from different subsamples of SDSS suggests that the Stripe 82 dataset is complete in mass down to a limit of \ms$\sim$8.5 for galaxies of all types (and likely higher for blue, star-forming galaxies which have lower M/L ratios than red sequence galaxies and thus are complete to lower masses in a given luminosity-limited sample). Hence the Stripe 82 data and ROLES may be fairly compared down to the ROLES mass limit of \ms$=$8.5. The blue and magenta dashed histograms show the Stripe 82 data selecting only blue cloud galaxies with SFRs$>$0.1\msunyr~and 0.3\msunyr, respectively. 

The mass function of z$\sim$1 star-forming galaxies from ROLES$+$FORS2 spectroscopy shows amazing agreement with the dashed blue histogram from SDSS local data. This histogram uses a SFR limit of 0.1\msunyr~which is a factor of 3 lower than the 0.3\msunyr~limit used at z$\sim$1. Arguably this is the fairest comparison since the average SFR density has decreased by a factor of $\sim$3 over this redshift range. This limit of 0.1\msunyr~locally also corresponds to the limited required for the SFRD to have converged \citep{Gilbank:2009nx}.  In order to quantify the difference between the mass functions at the two epochs, we construct the cumulative sum of each, i.e., $S_{z}=\sum_{m=11}^{m>M} \phi_m$ where $S_z$ is the sum at redshift z of the individual galaxy's number densities, $\phi_m$ for galaxies more massive than $M$, up to a maximum mass of \ms$=$11. This latter cut is placed so as to not be unduly affected by the small number of very massive galaxies in our sample. Examining the ratio of these cumulative mass functions at the two epochs, $S_{1.0}/S_{0.1}$ shows that the z$=$1.0 star-forming mass function is everywhere between  $\approx$6\% and $\approx$35\% higher than the z$=$0.1 star-forming mass function.\footnote{If instead of the 0.1\msunyr~SFR threshold (blue histogram) we adopt the 0.3\msunyr~SFR (magenta histogram), the difference grows steadily from $\approx$25\% above \ms=9.5 to $\approx$65\% at \ms$=$8.5.} Recently, \citet{Peng:2010rr} presented an elegant, empirical picture of galaxy evolution, one of the key ingredients of which is the apparent constancy of the mass function of star-forming galaxies over this redshift range and beyond. We have now measured this observationally over a much wider mass range than previously studied. The drop in the mass function of star-forming galaxies from z$=$1.0 to z$=$0.1 confirms that galaxies must be leaving the star-forming sequence/blue cloud, perhaps at an even faster rate than that assumed by \citet{Peng:2010rr}.

\section{Comparison with models}

In this section, the observed SSFR-mass relation is compared first with a simple toy model to parameterise galaxies' star formation histories as a function of their baryonic mass, and secondly with a recent state-of-the-art semi-analytical model of galaxy formation ({\sc galform}). The {\sc galform} predictions will also be confronted with additional observational results taken from Paper 2.

\subsection{Tau models}
\label{sec:tau}
One useful way to construct toy models for the star-formation histories of galaxies (e..g, \citealt{Savaglio:2005hp}, \citealt{Noeske:2007qa} ) is to adopt a closed-box model and make the instantaneous recycling approximation. If we relate the SFR, $\Psi$, to the mass of gas, $M_g$, via some star-formation efficiency factor, $\epsilon$, 
\begin{equation}
\Psi = \epsilon M_g,
\end{equation}
and assume a fraction, $R$, of the stellar mass formed is instantly returned to the ISM, this leads to an exponentially declining SFR (e.g., \citealt{Charlot:1991uq,Bruzual-A.:1993fk}\footnote{Such declining SFRs have also been historically referred to as ``$\mu$ models", where $\mu_{\mathrm SFR} = 1 - \exp (-1\,{\mathrm Gyr}/\tau)$}) with an e-folding time of $\tau$ where $\tau = 1/[\epsilon(1-R)]$, 

\begin{equation}
\Psi (M_b,z) = \Psi(M_b,z_f) \exp(-T/\tau),
\end{equation}
if we write $\Psi$ as a function of the baryonic mass, $M_b$, of the galaxy (initially all gaseous, i.e., $M_b=M_g$ at $z=z_f$). $z_f$ is the formation redshift of the stars.
\begin{equation}
T=t(z)-t(z_f)
\label{eqn:tz}
\end{equation}
where $t(z)$ is the cosmic time at which the galaxy is observed, and $t(z_f)$ is the cosmic time at $z_f$. Then,
\begin{equation}
\Psi(M_b,z_f) = \epsilon M_b = \frac{M_b}{\tau (1-R)}.
\end{equation}
The above, together with the relation
\begin{equation}
M_\star = \int^{t(z)}_{t(z_f)} \Psi(t)dt,
\end{equation}
which gives
\begin{equation}
M_\star(M_b,z) = \frac{M_b}{1-R}[1- \exp (-T/\tau)],
\end{equation}
specify $M_\star$ and $\Psi$ in terms of the baryonic mass, e-folding time, and formation redshift of the galaxy's stellar population. $R$ may be estimated from stellar population theory: $R=$0.56 for our BG03 IMF (e.g., \citealt{Hopkins:2006bv}).

\citet{Noeske:2007qa} describe a model of `staged galaxy formation' where they parameterise $\tau$ and $z_f$ as power laws of the baryonic mass, 
\begin{equation}
\tau(M_b) = c_\alpha M_b^\alpha
\end{equation}
 (a similar model was proposed by \citealt{Savaglio:2005hp} to explain the high redshift mass--metallicity relation), and
\begin{equation}
1+z_f(M_b) = c_\beta M_b^\beta.
\end{equation}

This methodology is applied to the data in Fig.~\ref{fig:ssfrmass} as a convenient way to parameterise the behaviour of the star-formation properties of galaxies. Simultaneous fits to the mean values (large symbols) at the two epochs are performed using {\sc mpfit}\footnote{see {\tt http://purl.com/net/mpfit}} in IDL \citep{Markwardt:2009bs}, allowing the parameters $\alpha$, $c_\alpha$, $\beta$, and $c_\beta$ to float free. The best-fit parameters are indicated on Fig.~\ref{fig:ssfrmass}, with the best fit curves overplotted as green solid lines for the z$=$1 data and green dashed curves at z$=$0.1. The best fit exponents in the (preferred) dataset from the right panel have values of $\alpha=-1.0$ and $\beta=0.3$, the same as those proposed by \citet{Noeske:2007qa}, albeit with different normalisations in order to fit the lower overall normalisation of the z$\sim$1 data. Due to the high mass limit in the DEEP2/AEGIS data, \citet{Noeske:2007qa} could only measure the average SSFR for the highest masses, (\ms$\sim$11), where they were $>$95\% complete. The low-mass end was an approximate by-eye fit to the bulk of the data points. Interestingly, if we combine our highly complete ($>$80\%) low-mass data with the DEEP2/AEGIS data a very different dependence is found (See \S\ref{sec:ssfrmass2}).

With the values fitted in Fig.~\ref{fig:ssfrmass}, the typical $z_f$ for a galaxy of \ms$=$[9, 11] at z$=$1 would be [1.7, 4.0] with $\tau=$[34, 4] Gyr; or for galaxies with these masses at z$=$0.1, the corresponding values would be $z_f=$[1.1, 3.7] and $\tau=$[64, 5] Gyr.

So, the upturn seen in the z$\sim$1 SSFR--mass relation towards lower masses can be explained by a non-zero $\beta$, i.e., the formation redshift was more recent for lower mass galaxies. Although a linear fit ($\beta=0$) in log--log space is permitted (\S\ref{sec:ssfrmass}), the higher mass points all lie systematically above the linear fit (Fig.~\ref{fig:ssfrmass}, right panel) and lower mass points systematically below, favouring the current parameterisation. With only the higher mass data used in other spectroscopic surveys, it is unlikely that this upturn could be detected.

Consider the implications for the mass function of star-forming galaxies (\S\ref{sec:mf}). In this toy model, individual star-forming galaxies are increasing in stellar mass: a typical \ms$=$9 galaxy at z$=$1 would increase its stellar mass by $\approx$0.5 dex by z$=$0.1 and a \ms$=$11 galaxy at z$=$1 by $\approx$0.1 dex. Some fraction of galaxies must also have their star-formation terminated and leave the blue cloud \citep{Arnouts:2007dz,Faber:2007sy}. These effects would act so as to shift the exponential cut-off of the mass function towards higher masses (and to shift the low mass tail toward higher masses by a larger amount), and to decrease the overall normalisation of the mass function, respectively. Therefore processes must be occurring in order to balance these effects, preserving the apparent constancy of the mass function \citep{Peng:2010rr}.

\subsection{The {\sc galform} model}

In order to gain some insight into the processes which might be responsible for the above results, we compare the observations with a recent semi-analytic model of galaxy formation, the {\sc galform} model of \citet{Bower:2006vw}, implemented within the Millennium $N$-body simulation. This is a version of the Durham semi-analytic model which includes feedback from AGN to quench cooling within massive haloes. 

Firstly we add additional data in order to test the model over a wider redshift baseline (\S\ref{sec:adddata}), and recast the SSFR--mass results in a manner more directly related to the predictions made by the model (\S\ref{sec:timescale}).

\subsubsection{Additional data, z$=$2}
\label{sec:adddata}

To the  $z=1$ and $z=0.1$ data shown in Fig.~\ref{fig:timescale} (blue and black points, respectively), we add higher redshift data at $z\sim2$ from two BX galaxy surveys (Sawicki \& Thompson 2006; and Reddy \& Steidel 2009, red squares and circles, respectively). These $z \sim 2$ samples are used to extend the redshift baseline of our comparisons.

It is important to note that the two stellar mass functions at  $z \sim 2$ are derived in a different way from those at lower redshifts.  The Sawicki (2010) MF is obtained from the BX galaxy LF using an empirical $M_{1700}$-stellar mass relation. This relation has been derived from a small but deep sample of $z \sim 2$ galaxies in the Hubble Deep Field for which stellar masses were estimated from multi-wavelength ($B$ through $H$) SED fitting, and which takes the form $M_\star = 0.68 M_{1700} - 0.46$ \citep{Sawicki:2011aa}.  Sawicki (2011) use it to infer a stellar mass function from a revised version of the Sawicki \& Thompson (2006) $z \sim 2$ UV LF that has been updated to take into account luminosity-dependent dust effects.  We follow the same approach but apply the mass-luminosity relation to the original LF (Sawicki \& Thompson 2006). This LF has a somewhat shallower faint-end slope than the updated LF; together with the Reddy \& Steidel (2009) result (below), which comes from a steeply-rising z$\sim$2 LF, the two results span a conservatively wide plausible range of z$\sim$2 possibilities. At their faint/low-mass end, Reddy \& Steidel (2009) follow a similar approach and an earlier version of the same HDF-based mass calibration. Specifically, at the low-luminosity end they correct their BX galaxy LF using their prescription for magnitude-dependent dust and then apply the Sawicki et al.\ (2007; see also Sawicki 2011) SFR-stellar mass relation to arrive at the low-mass end of the MF; at the bright/massive end, they use stellar masses obtained from SED fitting for individual galaxies in their large sample of BX galaxies.

The studies at $z\sim2$ have several key differences from our study at lower redshift.  The principal of these is that stellar masses are not derived for individual galaxies, as is the case at lower redshifts, but rather stellar mass densities are calculated from UV LFs and empirical stellar mass - $M_{1700}$ (or stellar-mass - SFR) relations --- an approach that can propagate significant scatter ($\sim 0.3$ dex) into the results.  Additionally, these conversions are based on the assumption of a single, constant star formation history in the SED fitting. Finally, it should be kept in mind that both the $z \sim 2$ samples rely on essentially the same stellar mass - $M_{1700}$ relation and the chief differences in the results stem from differences in their somewhat different UV luminosity functions.

\subsubsection{Timescale for growth}
\label{sec:timescale}
\begin{figure}
	{\centering
	\includegraphics[width=85mm,angle=0]{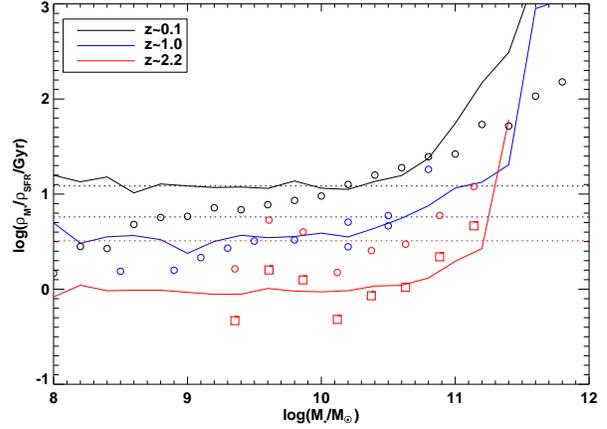}
	\caption{The timescale defined by $\rho_{M_\star}/\rho_{SFR}$ as a function of stellar mass for {\it all} galaxies rather than just the {\it star-forming} sample. Red symbols show z$\sim$1 data taken the SDSS Stripe 82 \citep{Gilbank:2009nx}; z$\sim$1 data (blue symbols) are taken from ROLES (SFRD) with the stellar mass function from \citet{Drory:2005ts} for the same field; red symbols are z$\sim$2 data taken from \citet{Sawicki:2006xw} (open red circles, SFRD) and \citet{Reddy:2009wc} (open red squares, SFRD) with the stellar mass functions taken from \citet{Drory:2005ts}. Dotted horizontal lines show the Hubble time at the corresponding redshift, colour-coded as above. Filled lines are {\sc galform} model predictions. See text for details.}
	 \label{fig:timescale}
}
\end{figure}

As shown in Eqn.~\ref{eqn:invssfr}, the inverse of the SSFR defines a characteristic timescale. Similarly, the inverse of the volume-averaged SSFR (Eqn.~\ref{eqn:r_v}) defines a timescale for the volume-averaged population ($\rho_{M_\star}/\rho_{SFR}$). \citet{Bower:2006vw} studied the dimensionless version of this quantity which they denote as $R=\rho_{M_\star}/\rho_{SFR}t_H(z)$, the ratio of the past to present SFR; where $t_H(z)$ is the age of the Universe at redshift $z$. The {\sc galform}  predictions from the \citet{Bower:2006vw} model are plotted in Fig.~\ref{fig:timescale}. Our results are overplotted. These results are closely related to the inverse of SSFR plotted in Fig.~\ref{fig:ssfrmass}. However there are some important differences. Note that since the \citet{Bower:2006vw} prediction is for {\it all} galaxies rather than the star-forming subsample as we have defined it, our estimate of $\rho_{M_\star}/\rho_{SFR}$ needs to include {\it all} galaxies. The change in $\rho_{SFR}$ moving from a star-forming sample to galaxies of all types should be minimal. Locally, \citet{Salim:2007rv} find that 95\% of the SFRD is in blue galaxies (as we have defined our star-forming sequence), and we discussed in \S\ref{sec:ssfrmass} the difference between using all z$\sim$1 galaxies with non-zero SFRs versus those selected to be in the blue cloud. In order to recalculate $\rho_{M_\star}$ for galaxies of all types, instead of using our measurement of the MF for only star-forming galaxies, we use the mass function of \citet{Drory:2005ts}. The difference between this total mass function compared with the star-forming mass function is $\approx$0.3 dex (although compatible within our broad uncertainties) at the high mass end, where we posses measurements of the SFR (\ms$\lsim$11), and smaller than this towards lower masses. The zCOSMOS-measured spectroscopic total mass function \citep{Pozzetti:2009qv} agrees well with the \citet{Drory:2005ts} measurement where they overlap at higher masses. Results at lower masses, where the photo-z estimate of the total mass function is not tested, should be regarded as the largest possible systematic error in the comparison with the models. 

Another, more minor, difference between Figs.~\ref{fig:ssfrmass}  and \ref{fig:timescale} is the distinction between the quantities defined in Eqns.~\ref{eqn:r_g} and \ref{eqn:r_v}. For these data, this difference is smaller than the random errors in a given mass bin. Note that the values in Fig.~\ref{fig:ssfrmass} have been weighted by Vmax and the completeness of the surveys.

This plot may be compared with the dimensionless ratio plotted in fig.~7 of \citet{Bower:2006vw}. The dimensionless quantities may simply be recovered by subtracting the log of the Hubble time, plotted as the dotted horizontal line, from the log of the timescale plotted as the curve. The main result of this comparison is the point at which galaxies of different masses make the transition from significant or increasing modes of star-formation to quiescent or declining SFRs. This is the mass at which the star-formation timescale at a given redshift equals the Hubble time at that redshift, i.e., the point at which the curve crosses the corresponding horizontal line. In the data, this transition mass occurs at \ms$\sim$11 at z$=$2 but at \ms$\approx$10.5 for z$=$1.0 and \ms$\approx$10.2 at z$=$1. This is one manifestation of cosmic downsizing in the SSFR: more massive galaxies have transitioned earlier from active to quiescent star-formation, relative to lower mass galaxies. This same trend is seen in the data. Indeed, the models and the data seem to agree best around the value of this transition mass. At z$=$0.1, at higher and lower masses, the model predicts somewhat higher values of the past-average to present SFR. We will return to this point below. Also, towards lower masses at z$=$1.0, the model is consistent with the individual measurement uncertainties\footnote{Error bars are not plotted for clarity, but these may be approximately gauged from the mass function errors in Fig.~\ref{fig:mf} and the SFRD errors in Fig.~\ref{fig:sfrd2}.} but lies systematically somewhat above the data. This means that the model does not reproduce the upturn at low masses in Fig.~\ref{fig:ssfrmass} which is driving our fit of a Tau model with non-zero $\beta$. At low masses, almost all galaxies should be star-forming at the level we can measure and so differences due to the ROLES data only selecting star-forming galaxies and {\sc galform} selecting all galaxies should be negligible. It is also worth noting that feedback from AGN is negligible in the models for low mass galaxies and so, whereas these effects could both play a part in the disagreement at the high mass end, the disagreement at the low mass end is pointing to inadequacies in the modelled physics for low mass galaxies.  It is beyond the scope of this paper to examine in detail what processes might be responsible, but we note that effects such as over-quenching for satellite galaxies (e.g., \citealt{Gilbank:2008pi}) may be at least partly responsible, and we point modellers to the current low mass data as an important test of galaxy formation theories.

\subsubsection{The SFRD as a function of stellar mass}
\begin{figure*}
	{\centering
	\includegraphics[width=85mm,angle=0]{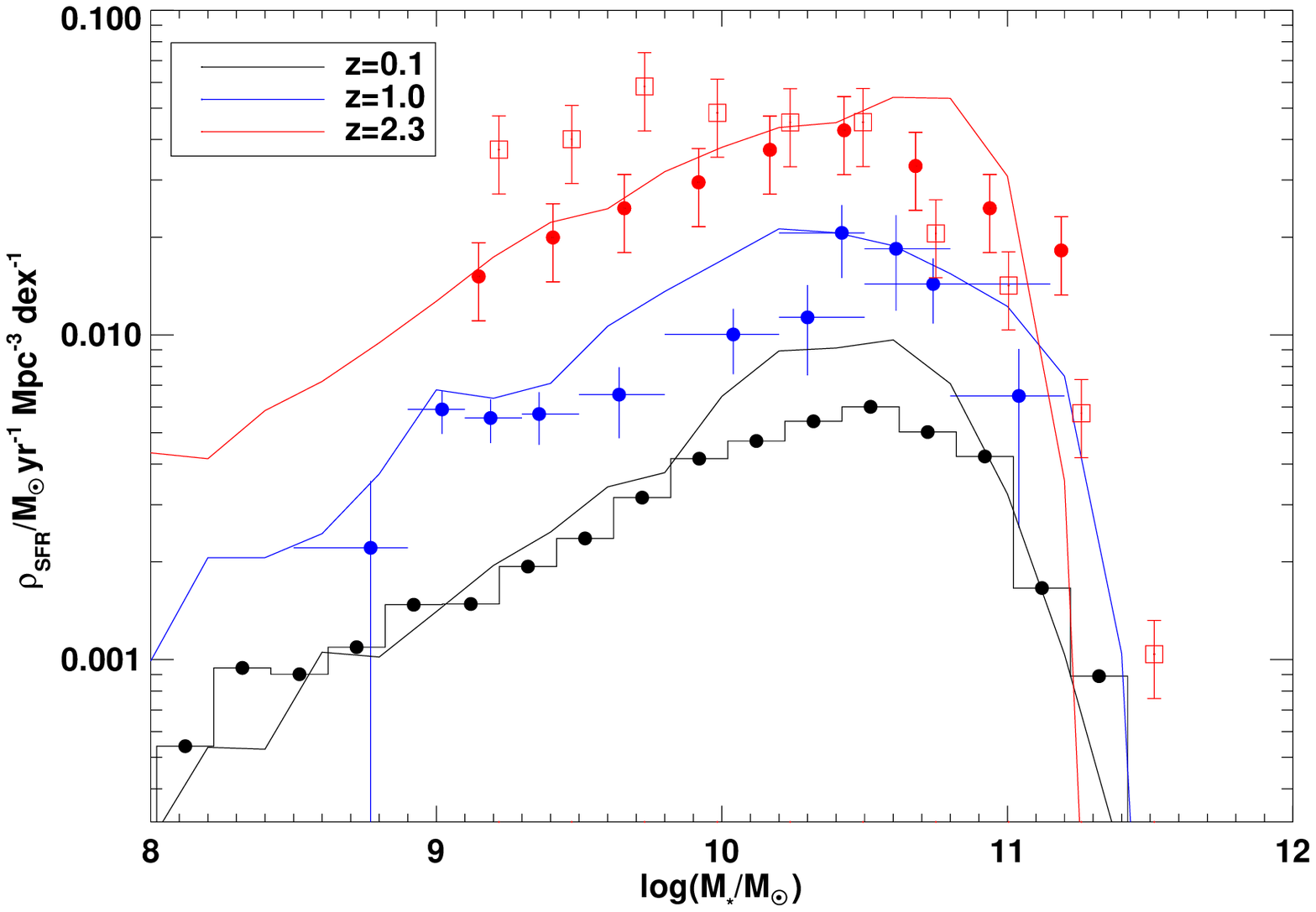}
	\includegraphics[width=85mm,angle=0]{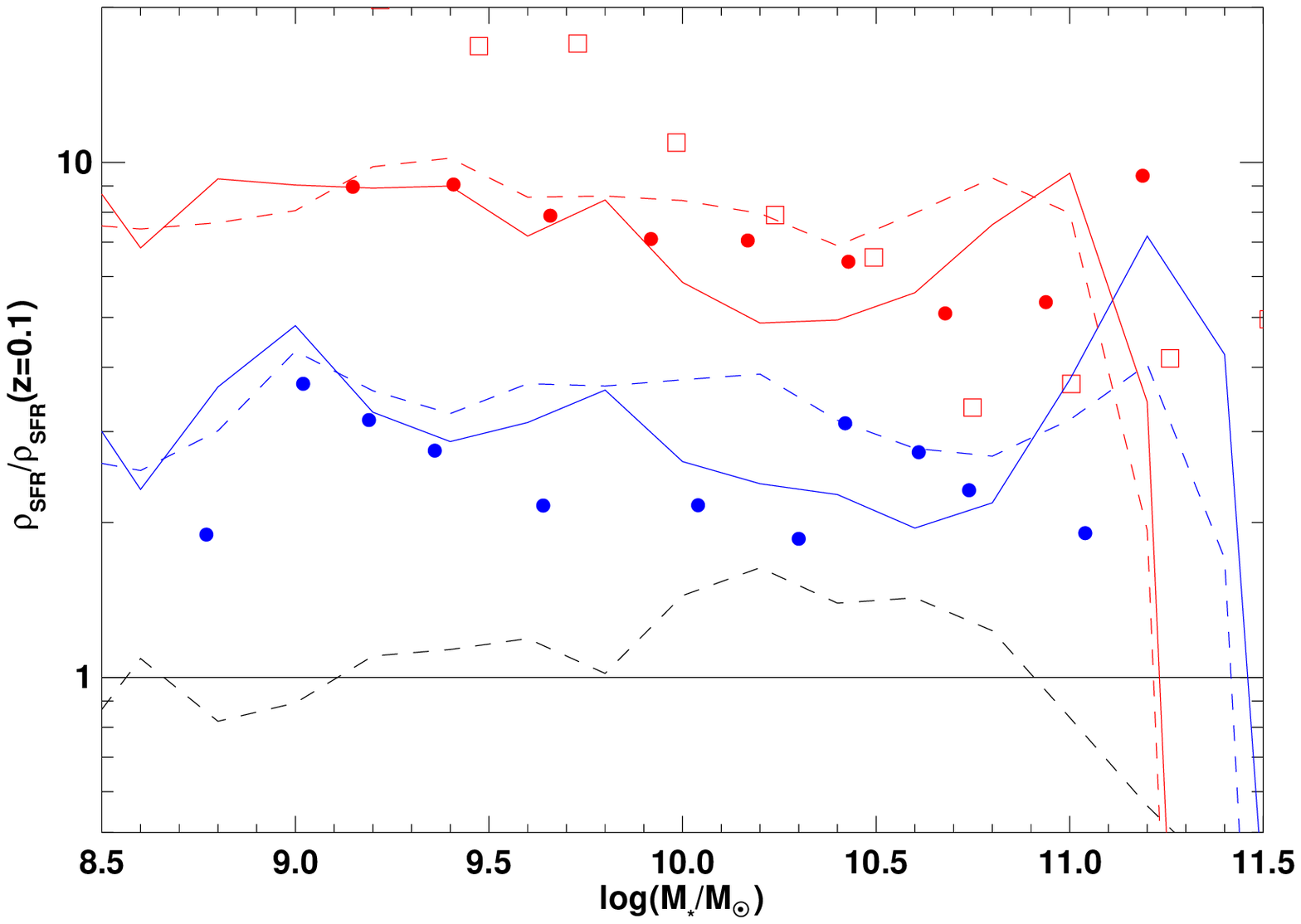}
	\caption{Evolution of the SFRD as a function of stellar mass. Left panel shows observed SFRD at z$\sim$0.1 (black histogram), z$\sim$1 from \oii~(blue filled circles with error bars), and z$\sim$2 (red filled circles \citep{Sawicki:2006xw}, open red squares \citep{Reddy:2009wc}). Solid curves are model predictions from galform where the redshifts are indicated by the same colours as the data. Right hand panel shows the same information, plotted as the ratio of the SFRD at the redshift of interest divided by the z$=$0.1 SFRD. Solid curves show the model predictions relative to the z$=$0.1 {\it model prediction}, and the dashed curves show the predictions dividing by the z$=$0.1 {\it data}.  Towards lower masses, the steeper gradient in the ROLES data reflects the low mass upturn seen in earlier plots. See text for details.
	}
	 \label{fig:sfrd2}
}
\end{figure*}

Another test for the {\sc galform} model is the evolution of the SFRD as a function of stellar mass. These were calculated for z$\sim$0.1 in \citet{Gilbank:2009nx} and z$\sim$1 in Paper 2 using the standard 1/Vmax technique as described in those papers. To these, the data from \citet{Sawicki:2006xw} and \citet{Reddy:2009wc} are again added, converting to SFRD--mass as described above. Fig.~\ref{fig:sfrd2} (left panel) shows the SFRD, $\rho_{SFR}$ as a function of stellar mass in three different redshift bins (left panel). The different coloured points refer to data at the different redshifts (black: z$\sim$0.1, blue: z$\sim$1, red: z$\sim$2). Again, the lowest redshift data (black histogram) are taken from Stripe 82 data, where $\rho_{SFR}$ is calculated as described in \citet{Gilbank:2009nx}. The z$\sim$1 $\rho_{SFR}$ are taken from the (empirically-corrected) \oii~SFRD in Paper 2. The z$\sim$2 $\rho_{SFR}$ data are taken from the LBG samples of \citet{Sawicki:2006xw} (filled red circles) and \citet{Reddy:2009wc} (open red squares). The overplotted coloured curves show the prediction of the {\sc galform} model. Errors are Poisson errors from the number of objects and do not include uncertainties associated with assuming different star-formation histories, etc. (but they do encapsulate the measured scatter in the adopted SFR--mass relation).

The right panel shows the same information, but now plotted as the ratio between the SFRD at the redshift of interest and that at z$=$0.1, $\rho_{SFR}(z=0.1)$. Solid curves show the model predictions relative to the z$=$0.1 {\it model prediction}, and the dashed curves show the predictions dividing by the z$=$0.1 {\it data}. 

The overall normalisation of the {\sc galform} model SFRD at all three epochs is reasonably well-matched to the observations. The best agreement occurs for the lowest masses at $z$=0.1 and z$=$1.0 where the ROLES data are best measured. A larger disagreement occurs around \ms$\approx$10--10.5, where the model overpredicts the data at both epochs. Interestingly this is around the same place as the best agreement occurs in the timescale plot (Fig.~\ref{fig:timescale}). Since the timescale is simply the stellar mass function divided by the SFRD, this implies a corresponding systematic difference in the {\sc galform} mass function. Comparing  the {\sc galform} mass function with that of the \citet{Baldry:2008sj} total stellar mass function (plotted in Fig.~\ref{fig:mf}), the former lies systematically above the latter at the highest and lowest masses \ms$\lsim$10.5 and \ms$\gsim$11.0, but within the 1$\sigma$ errors of each point. Interestingly, both the data and model showed a downsizing trend in the star-formation timescale (as described above). However, we claimed in Paper 2 that no downsizing trend in SFRD was seen in our data, and that the SFRD at z$=$0.1 could be fit by a renormalised version of the local SFRD. The {\sc galform} predictions are in good agreement with our data, where the effect of downsizing should be most obvious. Our highest mass datapoint is only marginally discrepant ($<$2$\sigma$). Differences are most easily seen in the ratio plots in the right hand panel of Fig.~\ref{fig:sfrd2}. On this plot error bars have been omitted for clarity, but may be gauged from the left panel. The ratio of model SFRDs at z$=$1.0 to z$=$0.1 (solid blue line) is very nearly independent of mass, and it is only at \ms$>$11.2 where a significant difference is seen. For the z$=$2 data, both observational results indicate the peak of the SFRD lies towards lower masses than that suggested by the {\sc galform} model. Indeed, examining the right panel of Fig.~\ref{fig:sfrd2}, the shape of the SFRD--mass relation seems to have changed remarkably little since z$\sim$0.1. 

\begin{figure}
	{\centering
	\includegraphics[width=85mm,angle=0]{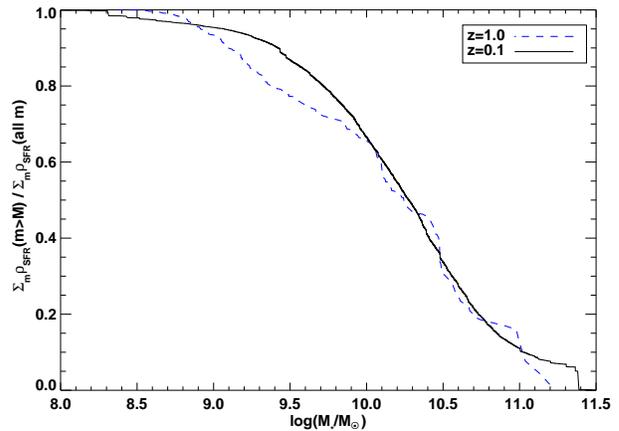}
	\caption{Fractional cumulative SFRD as a function of stellar mass for the two spectroscopic surveys of ROLES$+$FORS2 (blue dashed line) and SDSS Stripe 82 (black solid line). At high masses, the two distributions are remarkably similar. More than 70\% of the SFRD occurs in galaxies more massive than \ms$>$10 at both epochs. See text for discussion. 
	}
	 \label{fig:cumsfrd}
}
\end{figure}

In order to more closely look for differences between the SFRD--mass distributions at the different epochs, Fig.~\ref{fig:cumsfrd} shows the fractional cumulative SFRD in galaxies more massive than a given stellar mass. The distributions are only plotted for the spectroscopic (z$=$0.1 and z$=$1.0) data, where the measurements are most secure and uniform. At high masses (\ms$>$10), the two distributions are very similar. Indeed, almost 70\% of the total SFRD at each epoch occurs above this mass limit. At \ms$\sim$9.5, the z$=$1 distribution drops below that of the local value. This may be due to inadequate sampling of galaxies in this mass range, as this is the point around which the FORS2 data end and the ROLES data begin. At the low mass  limits of the data, the ROLES distribution begins to exceed the local measurement, and the gradient in the former is much steeper. This reflects the upturns we see in Figs.~\ref{fig:ssfrmass} \& \ref{fig:mf}. If the deficit around \ms$\sim$9.5 is due to a paucity of these mass galaxies in our sample, then the excess measured here is likely even higher than 5\%. This shows that high mass galaxies (\ms$\gsim$10) appear to have evolved equally as a function of stellar mass between z$=$0.1 and z$=$1.0, in SSFR (and its inverse, timescale) and SFRD, whereas low mass galaxies have evolved more (as shown by Figs.~\ref{fig:ssfrmass} and \ref{fig:cumsfrd}). The strength of any statement on this differential evolution is limited by the size of the current high mass sample. 

\subsection{Other semi-analytic model comparisons in the literature}

Other works have compared their observations with different semi-analytic models applied to the Millennium simulation. \citet{Elbaz:2007rt} used UV$+$24$\mu$m SFRs and spectroscopic redshifts in the GOODS-N and GOODS-S (CDFS) fields to test the models of \citet{Croton:2006uk}. They found an increase in the normalisation of the SFR--M$_\star$ relation between z$\sim$0 and z$\sim$1 of a factor of 6 in the data, and an increase of only a factor of 2 predicted by the models over the same range. \citet{Damen:2009it} used UV$+$24$\mu$m data with FIREWORKS' photometric redshifts in the CDFS to compare with the models of \citet{De-Lucia:2007xw}. They found that the logarithmic increase of the SSFR with redshift is nearly independent of mass, for the higher mass bins which could be tested (down to a limit of \ms$\sim$9.5 at z$=$0.9)\footnote{Note that although we use the same FIREWORKS data, we do not require such a high significance limit on the $K$-band magnitude as \citet{Damen:2009it}, so we push the available photometry deeper by requiring an emission line to obtain a redshift, rather than fitting high-precision photometry to obtain a photometric redshift.}. Over the same mass range in the SDSS and FORS2 data (Fig.~\ref{fig:ssfrmass}), our data appear consistent with the same result. They found that the same trend was seen in the \citet{De-Lucia:2007xw} models, as it also appears to be in {\sc galform} from the parallel nature of the curves in Fig.~\ref{fig:timescale}. 

\citet{Damen:2009it} also used the dimensionless growth rate ($GR_{sf}=(\Psi/M_\star)t_H(z)$), i.e., closely related to the inverse of the star formation timescale divided by the age of the Universe at the redshift of interest (and the inverse of \citealt{Bower:2006vw}'s $R$), as the observational quantity to compare with semi-analytic models of galaxy formation. The predictions were made by \citet{Guo:2008kl} which are based on the model of \citet{De-Lucia:2007xw}. As with the \citet{Bower:2006vw} model, these semi-analytic prescriptions are applied to the Millennium simulation. \citet{Damen:2009it} found that the growth rate was in good agreement with the models at z$\sim$0, but that the growth rate in the observations increased much faster than those of the models. The \citet{De-Lucia:2007xw} model seems to predict a much flatter evolution in the growth rate/timescale than the \citet{Bower:2006vw} model over the redshift range considered here (compare fig.~1 of \citealt{Guo:2008kl} with  fig. 7 of \citealt{Bower:2006vw} model). Thus, although both semi-analytic models qualitatively reproduce the downsizing trend in cosmic star-formation, they both fail to reproduce important details: the \citet{De-Lucia:2007xw} predicts too slow a rate of the growth of stellar mass due to star-formation, although the values for different galaxy masses at z$=$0 are quite accurately predicted; whereas the \citep{Bower:2006vw} model predicts approximately the correct growth rate for all but the most massive galaxies (\ms$\gsim$11), underpredicting the local SFRD in high mass galaxies. Conceivably these shortcomings could be due to similar factors, however a detailed investigation into the semi-analytic models is beyond the scope of this paper. 

\citet{Damen:2009it} and others (e.g., \citet{Zheng:2007vl}) have claimed that the evolution in growth rate, or equivalently SSFR, is the same for galaxies of all masses. If our limiting mass was higher (more like \ms$\gsim$9.5) we would have reached similar results. Our lower mass sample identifies an upturn in average SSFR at lower masses which we interpret as evidence for a later formation epoch for lower mass galaxies in our toy-modelling.

\section{Possible systematic errors}
\label{sec:discuss}

\subsection{SFR calibration}

Our measurements are most robust for low mass galaxies, for the reasons mentioned above. 
More uncertainty exists when results for high mass galaxies are included. The high mass data used comes from public spectroscopy of the CDFS. This is a relatively small field (which is especially problematic for the rarer, high mass galaxies) and these data have a sampling completeness of $\approx$50\% (and an overall completeness, including the redshift success rate, of $\approx$30\%). However, the agreement between the CDFS data and the larger zCOSMOS survey converted to the same empirically-corrected \oii-SFR tracer gives reassurance that cosmic variance does not have a significant impact on these results (Appendix \ref{sec:ssfrmass2}). Another concern is how well the empirically-corrected \oii-luminosity traces the total (i.e. extinction-corrected) SFR. The calibration is known to give good agreement between extinction-corrected \ha-SFRs at z$\sim$0.1 \citep{Gilbank:2009nx}, where the extinction has been calibrated either by the Balmer decrement or IR luminosity. The disagreement between the DEEP2/AEGIS SSFRs and the current dataset (see Appendix~\ref{sec:ssfrmass2}) at the high mass end might suggest that the empirically-corrected \oii-SFRs are missing extinguished star-formation detected in the 24$\mu$m observations (but note that the dicrepancy still persists for \oii-only SFRs); or recent suggestions that the rest-frame 12$\mu$m data overestimates the true SFR may be responsible (see Appendix~\ref{sec:ssfrmass2}).   

The validity of a mass-dependent correction for \oii-SFRs must be tested with an independent SFR indicator less-sensitive to extinction. It is an important open question \citep{Kennicutt:2009ng} whether extinction is more physically related to a galaxy's stellar mass or SFR (e.g., \citealt{Hopkins:2001db,Brinchmann:2004ct, Strazzullo:2010zm}). Recently \citet{Garn:2010bf} have presented compelling evidence from a large SDSS sample that suggests stellar mass is the more fundamental, supporting our mass-dependent empirical correction for the effects of dust. In addition, \oii~luminosity is also sensitive to the effect of metallicity. As we discussed in paper 2, it is an open question how the effects of evolution in metallicty and dust may change our mass-dependent correction to the \oii-SFR between that calibrated locally and that at higher redshift. It is possible that the evolutionary effects may in fact cancel out, since at z$\sim$1 galactic metallicity is lower than locally (so \oii~luminosity increases for a given SFR), and dust extinction is likely higher \citep{Tresse:2007fu,Villar:2008bv} (so \oii~luminosity decreases for a given SFR). Indeed, \citet{Mannucci:2010kb} recently proposed a fundamental relation between stellar mass, metallicity and SFR which is observed to remain constant, due to a conspiracy in the evolution of these parameters, up to z$\sim$2.5. We estimate the likely size of these effects below.

For low mass galaxies, the situation is somewhat simplified as a) the dust extinction is extremely low, as we have shown above; b) theoretical models for the dependence of \oii~luminosity on SFR as a function of metallicity \citep{Kewley:2002du} give good agreement with empirical calibrations \citep{Gilbank:2009nx}. Eqn. 15 from \citet{Kewley:2002du} gives the correction to SFR which must be applied to \oii~luminosity as a function of metallicity. Taking the mass--metallicity (M--Z) relations from \citet{Savaglio:2005hp} for z$\sim$0.1 and z$\sim$0.7 (their eqns. 8 and 7 respectively), one can estimate the change in this theoretical correction between z$\sim$0.7 and z$\sim$0.1. For a \ms$\approx$9.0 galaxy (which is well sampled by the \citealt{Savaglio:2005hp} data at z$\sim$0.7), the evolution in this correction due to the evolution in metallicity is $\approx$5\%. This is not large enough to explain the observed upturn in the SSFR--mass relation for low mass galaxies. Towards higher masses, the metallicity difference (between z$\sim$0.7 and z$\sim$0.1, eqns. 7 and 8 of  \citealt{Savaglio:2005hp}) becomes smaller, in fact crossing at \ms$\approx$ 10.3\footnote{This may not be physical and may be due to the use of a linear fit to the z$\sim$0.7 data, and a polynomial fit to the z$\sim$0.1 data. Nevertheless the metallicity difference at these masses is still small.}, and so differential evolution in the M--Z relation cannot explain the upturn at \ms$\lsim$9 relative to \ms$\sim$10 galaxies. Furthermore, the evolution in average dust extinction between z$\sim$0.1 and z$\sim$1 appears to be mild, even for the more massive galaxies considered here (e.g., \citealt{Garn:2010fk}). Thus, the expectation is that the empirical mass-dependent correction for \oii-SFR should not change significantly by z$\sim$1. To properly test our empirical correction would require an analogue of our local test using Balmer decrement corrected \ha-SFRs at z$\sim$1, and such work is ongoing. 

\subsection{Sample selection}
\label{sec:sampsel}
Another possible consideration is how star-forming galaxies are selected. It is now well-established that galaxies show a bimodal distribution in colour out to at least z$\sim$1 \citep{Strateva:2001db,Baldry:2004wj,Bell:2004lb}, comprising a `blue cloud' dominated by star-forming galaxies, and a red-sequence dominated by passive galaxies with some minority of dust-reddened star-forming galaxies. A similar bimodality is also seen locally in the distribution of SFR or SSFR  (\citealt{Brinchmann:2004ct, Salim:2007rv}, McGee et al., MNRAS submitted), such that galaxies exhibit a relatively tight star-forming sequence (which may have scatter as low as 0.2 dex  \citep{Salim:2007rv} when plotted against stellar mass), and a broader population of galaxies with low (S)SFRs exhibiting little or no star-formation. However a tail exists down from the star-forming sequence such that it is necessary to go $\gsim$1 dex lower in SFR to encompass 95\% of galaxies\footnote{where galaxies with spectroscopic indications of AGN activity are excluded} (fig.~17,  \citealt{Brinchmann:2004ct}). Thus, however galaxies have their SFRs measured, be it emission lines or SED-fitting (such as \citealt{Salim:2007rv}, McGee et al., in prep), it is necessary to identify the star-forming sequence using either an overdensity selection in (S)SFR--mass space; or an approximate cut on (S)SFR; or, as we choose here, a colour cut to isolate the blue cloud. This rejects the tail of galaxies towards lower SSFRs and allows the bulk of the star-forming sequence to be isolated. This colour cut has the advantage when used with \oii-SFRs that it rejects objects for which the source of the emission is likely not star-formation, and it simplifies the comparison with other works where people have applied a colour selection. Throughout this work, we have applied a colour cut to exclude red galaxies (Fig.~\ref{fig:cmd}). This makes negligible difference to any of our results for dwarf galaxies  since only one galaxy out of 199 in ROLES is red by this criterion, and the low mass end in the SDSS data is unaffected by the colour cut (c.f. left and right panels of Fig.~\ref{fig:ssfrmass}). At the high mass end, the z$\sim$1 data move slightly toward lower SSFRs (but still within the 1$\sigma$ errors) when red galaxies are included in the computation of the mean (c.f. left and right panels of Fig.~\ref{fig:ssfrmass}), suggesting that LINER activity and/or dust-reddened star-formation has little impact on the FORS2 sample. At z$\sim$0.1, a significant lowering of the average SSFR in the SDSS Stripe82 data occurs when red galaxies are included in the mean. This is caused by the wide tail of low SSFR galaxies extending from the main star-forming sequence which can be seen as the difference between the two panels of Fig.~\ref{fig:ssfrmass}. Excluding red-sequence galaxies better traces the peak of the contours (i.e. the mode) of the star-forming sequence and is more directly comparable to the selection made at z$\sim$1, and so we favour this approach.

\section{Conclusions}
\label{sec:concl}
We have presented results from a spectroscopic sample of unprecedented depth at z$\sim$1, probing star-formation via \oii~in galaxies down to \ms$\sim$8.5. This is combined with an equally-deep comparison sample taken from Stripe 82 of the SDSS. We have, for the first time, measured the SSFR--mass relation and mass function of star-forming galaxies down to \ms$\sim$8.5 at z$\sim$1 using spectroscopy. 

The strength of the ROLES dataset is that it offers a highly-complete spectroscopic sample for such low mass galaxies. Measurements for these dwarf galaxies are particularly robust since they are numerous enough to provide useful statistics, and systematic uncertainties in SFRs due to extinction are negligible since they possess negligible dust, as confirmed by the lack of 24$\mu$m emission (Appendix \ref{sec:stack24}). 

Dwarf galaxies are the building blocks of larger galaxies in the hierarchical formation scenario. The combination of ROLES and Stripe82 data extends these observations at unprecedented depth to a wide redshift baseline, allowing evolution to be examined in detail for this important population of galaxies.

The mass function for star-forming galaxies has remained remarkably constant over this interval, dropping by $\approx$(6-35)\% from z$=$1.0 to 0.1. This confirms that galaxies are leaving the star-forming sequence/blue cloud.

The SSFR--mass relation decreases in normalisation towards the present day. Although at high masses this relationship evolves almost in parallel, the evolution is greater toward lower masses (\ms$\lsim$9.5). The evolution of this relation can be approximately modelled by the staged galaxy formation toy model of \citet{Noeske:2007qa} in which galaxies' SFRs decline exponentially with a timescale dependent on their baryonic mass ($\tau \propto M_b^{-1}$), and also the formation redshift depends on the galaxy baryonic mass ($1+z_f \propto M_b^{0.3}$). This low mass upturn can also be seen in SFRD--mass by looking at differences in the cumulative distributions. Above \ms$>$10, the z$=$0.1 and z$=$1.0 SFRDs are extremely similar, both containing $\approx$70\% of the cosmic SFRD at this epoch above this mass. This upturn is subtle, and at the limit of our data, but we have investigated and shown that it is unlikely to be due to incompleteness, or evolution in the mass--metallicity relation or dust content of galaxies.

The \citet{Bower:2006vw} version of the {\sc galform} semi-analytic model of galaxy formation makes predictions for low mass galaxies which are in reasonable agreement with the measured SSFR--mass and SFRD--mass relations. At the lowest redshifts, the highest and lowest mass model galaxies exhibit too high a ratio of past-to-present SFR, possibly suggesting that star-formation has been quenched too early in the model.  At z$\sim$1, the measurements are most robust at the low mass end, and this suggests that the {\sc galform} model does not reproduce the upturn seen in the observational SSFR--mass relation at these low masses. One possible cause may be the too efficient termination of star formation in satellite galaxies (``strangulation'').

Future work will extend these results to higher redshift, which is the interesting epoch for downsizing as hinted at by the preliminary z$=$2 results, using the ROLES' technique with the new generation of red-sensitive CCDs. Work is ongoing in parallel to investigate the reliability of the z$\sim$1 empirically-corrected \oii-SFRs using NIR-MOS.

\section*{Acknowledgments}
We thank the referee for concise and perceptive suggestions encouraging us to perform additional tests of our results. We thank K.~Noeske and C.~Maier for providing electronic versions of their results and for useful discussions regarding their data. We thank S.~Wuyts and the FIREWORKS team for making their excellent catalogue publicly available. This paper includes data gathered with the 6.5-m Magellan Telescopes located at Las Campanas Observatory, Chile.   Karl Glazebrook and I-H Li acknowledge financial support from Australian Research Council (ARC) Discovery Project DP0774469. Karl Glazebrook and Ivan Baldry acknowledge support from the David and Lucille Packard Foundation. This research was supported by an Early Researcher Award from the province of Ontario, and by an NSERC Discovery grant.

\appendix

\section{IMF conversions}
\label{sec:imf}
Different workers often assume different IMFs (each of which is invoked to fix various problems with the adoption of an earlier IMF, such as \citealt{Salpeter:1955zs}) when computing stellar masses and SFRs. These derived quantities, in principle, may have different dependancies on the IMF since they are sensitive to stellar populations with different lifetimes. Although, in detail, assumptions other than the IMF may play a significant role in the quantities derived (particularly for the stellar mass-fitting where the stellar population synthesis models used, the star-formation histories considered, and the choice of priors, etc. likely dominate the uncertainties, e.g., \citealt{Marchesini:2009fx}), one can attempt to correct for systematic offsets in a broad sense. Derivations from the literature for IMF transformations (e.g., \citealt{Bell:2003iq}, \citealt{Savaglio:2005hp}, \citealt{Baldry:2008sj}) are combined with conversions based on PEGASE.2 models as outlined in \citet{Gilbank:2009nx} to produce conversion factors from several of the most commonly-used IMFs (all the ones used in this paper: \citealt{Salpeter:1955zs, Kroupa:2001ea,Kennicutt:1983cz}) and the BG03 IMF. These are listed in Table~\ref{tab:imfs}. The SFR conversion factors are based on the relative luminosity of the \ha~line in scenarios with different IMFs, and are thus directly applicable to \oii~which is calibrated through the empirical relation with \ha.  Another distinction not always stated on the literature concerns which stellar objects are included in the estimate of mass. For example, ROLES and FORS2 derived stellar masses at z$\sim$1 (Paper 2) only include main sequence and giant stars and do not include the mass locked up in stellar remnants. Including or omitting remnants makes $\approx$0.1 dex difference (for the BG03 IMF), assuming solar metallciity and a population age of 10 Gyr, smaller than the random mass errors of 0.2dex \citep{Glazebrook:2004zr}. 

\begin{table}
\caption{Correction factors to convert stellar masses, M$_\star$, and SFRs, $\Psi$, to the BG03 IMF. Numbers give the values to be added to quantities in the original IMF to obtain the corresponding value for a BG03 IMF. For example, $\log M_{\star, \mathrm BG03} = \log M_{\star, \mathrm Kroupa01} -0.08$. These conversions have some dependence on star-formation history and metallicity. See text for details. 
\label{tab:imfs}}
\begin{tabular}{lcc}
\hline
IMF & $\log ({\mathrm M_{\star})_{corr}}$ & $\log(\Psi)_{corr}$ \\
\hline
\citet{Kroupa:2001ea} & -0.08 & -0.18 \\
\citet{Salpeter:1955zs} & +0.11 & -0.26 \\
\citet{Kennicutt:1983cz} & +0.04 & -0.37 \\
\hline
\end{tabular}
\end{table}

It is worth emphasising that the empirical mass-dependent correction to \oii~derived by \citet{Gilbank:2009nx} is based on stellar masses and SFRs assuming the Kroupa IMF. In order to apply eqn.~8 of  \citet{Gilbank:2009nx}, stellar masses and SFRs must first be converted to the Kroupa IMF and then back to the IMF of choice.

\section{Comparison with other \MakeLowercase{z}$\sim$1 \oii~spectroscopic surveys}

\subsection{Additional samples}
Additional z$\sim$1 data are taken from DEEP2 spectroscopy of the AEGIS field (\citealt{Noeske:2007tw}, kindly provided by K. Noeske). These data are restricted to the region covered by MIPS 24$\mu$m imaging and where the NIR photometry was deeper than $K=22$. Stellar masses were fitted by \citet{Bundy:2006lh}. We have restricted the redshift range of this data to that of ROLES (0.88$<$z$\le$1.15) and so the emission line estimates from DEEP2 all come from \oii. Unlike the above data, SFRs were not estimated purely from \oii~luminosity, but from the published data come from a combination of \oii$+$24$\mu$m luminosity where 24$\mu$m emission is significantly detected, and from \oii~otherwise. So, some care is needed when comparing these different indicators. 24$\mu$m SFRs were computed by template-fitting \citep{Le-Floch:2005bv,Chary:2001fu}; and emission line SFRs were measured by \citet{Weiner:2007tf}. We have converted the assumptions used by the DEEP2 team in computing the purely-\oii-SFRs to be the same as ours (i.e., they assume \oii/\ha$=$0.69 and an average extinction at \ha~of 1.30 mag independent of mass/magnitude; we adopt a nominal  \oii/\ha$=$0.50 and 1 mag of extinction at \ha, to which the empirical mass-dependent correction of \citealt{Gilbank:2009nx} can then be applied directly). However, we cannot separate out the contribution of the two components in the \oii$+$24$\mu$m SFRs.\footnote{Ideally, to correct \oii$+$24$\mu$m SFRs, one would like to apply the empirical correction of \citet{Gilbank:2009nx} (which corrects for mass-dependent extinction {\it and} metallicity trends) to the \oii, and then divide this by the mass-dependent extinction (eqn. 9 of \citealt{Gilbank:2009nx}) to remove the dust correction, to which one may add the 24$\mu$m flux which accounts directly for the extinguished component of star-formation .}
Thus, the DEEP2 data consist of \oii-SFRs using our mass-dependent correction (where no 24$\mu$m detection exists) or an uncorrected \oii$+$24$\mu$m SFR (where a significant 24$\mu$m detection exists). The DEEP2 SFRs are only calculated for blue galaxies \citep{Noeske:2007tw}, since the emission line luminosity for red galaxies primarily comes from AGN/LINER activity rather than star-formation. All measurements of SFR and stellar mass are converted to our adopted BG03 IMF (see below). 

As a check of the other two z$\sim$1 datasets, \oii~SFRs from the zCOSMOS survey (\citealt{Maier:2009sf}, kindly provided by C. Maier) are used. These cover a slightly different redshift range ($0.7<z\le0.9$) from that of the ROLES and DEEP2 data, and the effect of evolution over this range is examined in \S\ref{sec:ssfrmass2}. The \citet{Maier:2009sf} SFRs were calculated using the empirical $M_B$-corrected formula of \citet{Moustakas:2006wp}. This is a better estimate of the SFR than just using \oii~luminsoity, but still significantly underestimates the SFR at high stellar masses \citep{Gilbank:2009nx}. So, to correct for this and to measure SFR in a way consistent with the other samples, the $M_B$-corrected \oii~SFRs are multiplied by the ratio between the mass-dependent corrected \oii~SFR and the $M_B$-corrected SFRs at z$\sim$1 from fig.~15 of Paper 2\footnote{This is actually the ratio of the SFRDs but the data use exactly the same galaxies. In this way, the ratio gives a volume-weighted correction between the two SFR estimators.}. These corrections are $\approx$0.2-0.3 dex at \ms$\sim$11.

The above two datasets will next be compared with our data in the CDFS. To recap, all galaxies defined as `star-forming' in these datasets, have a spectroscopic redshift, significant non-zero \oii~emission, and lie in the blue cloud. In the case of a subsample of the DEEP2 data, some also possess significant 24$\mu$m emission (but may or may not additionally exhibit \oii~emission). 

\subsubsection{24$\mu$m data}
\label{sec:24data}
In order to fairly compare the combined \oii$+$24$\mu$m SFRs from \citet{Noeske:2007tw} with SFR estimated solely from empirically-corrected \oii, we turn to 24$\mu$m observations of our CDFS sample. The CDFS possesses deep 24$\mu$m MIPS observations from {\it Spitzer} (Dickinson et al., in prep.). MIPS 24$\mu$m fluxes are taken from the FIREWORKS catalogue for objects in the spectroscopic range of interest ($0.88<z\le1.15$). One of the main disadvantages of MIPS photometry is that the PSF is relatively large (6\arcsec) and so faint sources quickly become confused. To minimise this effect, we begin by just considering high mass sources, which will generally have higher SFRs and hence should be the brighter 24$\mu$m sources. To achieve this, objects are selected from just the FORS2 sample. In order to attempt to quantify the impact of confusion (which may still be problematic for these brighter sources), objects with significant IRAC band 1 flux (3.6$\mu$m, hereafter, [3.6]) are selected from the FIREWORKS catalogue. The [3.6] flux is often used as a prior when attempting to deblend the 24$\mu$m images, since typically every 24$\mu$m source is detected in [3.6] and the PSF of the latter  is much better than the former (1.6\arcsec). The [3.6] source list is cross-correlated with the 24$\mu$m list, and objects with more than one match with 6\arcsec~are  flagged as potentially blended. The 6\arcsec~radius means that two 24$\mu$m sources separated by this amount will touch at their half-maxima. This should only be regarded as a crude estimate of confusion since the FIREWORKS catalogue is $K$-band selected, and so a number of fainter [3.6] sources do not appear even in the deep $K$-band imaging, and hence some potential blends may not be identified. In addition, the presence of a second [3.6] source within the matching radius does not guarantee it will be a significant source of 24$\mu$m flux, and so cases flagged as `blended' may in fact suffer no contamination from a neighbour. Nevertheless, this method provides an approximate division between blended/unblended sources.

\subsection{Comparison of \oii~with 24$\mu$m SFRs}
\subsubsection{High mass galaxies}

\begin{figure}
	{\centering
	\includegraphics[width=105mm,angle=0]{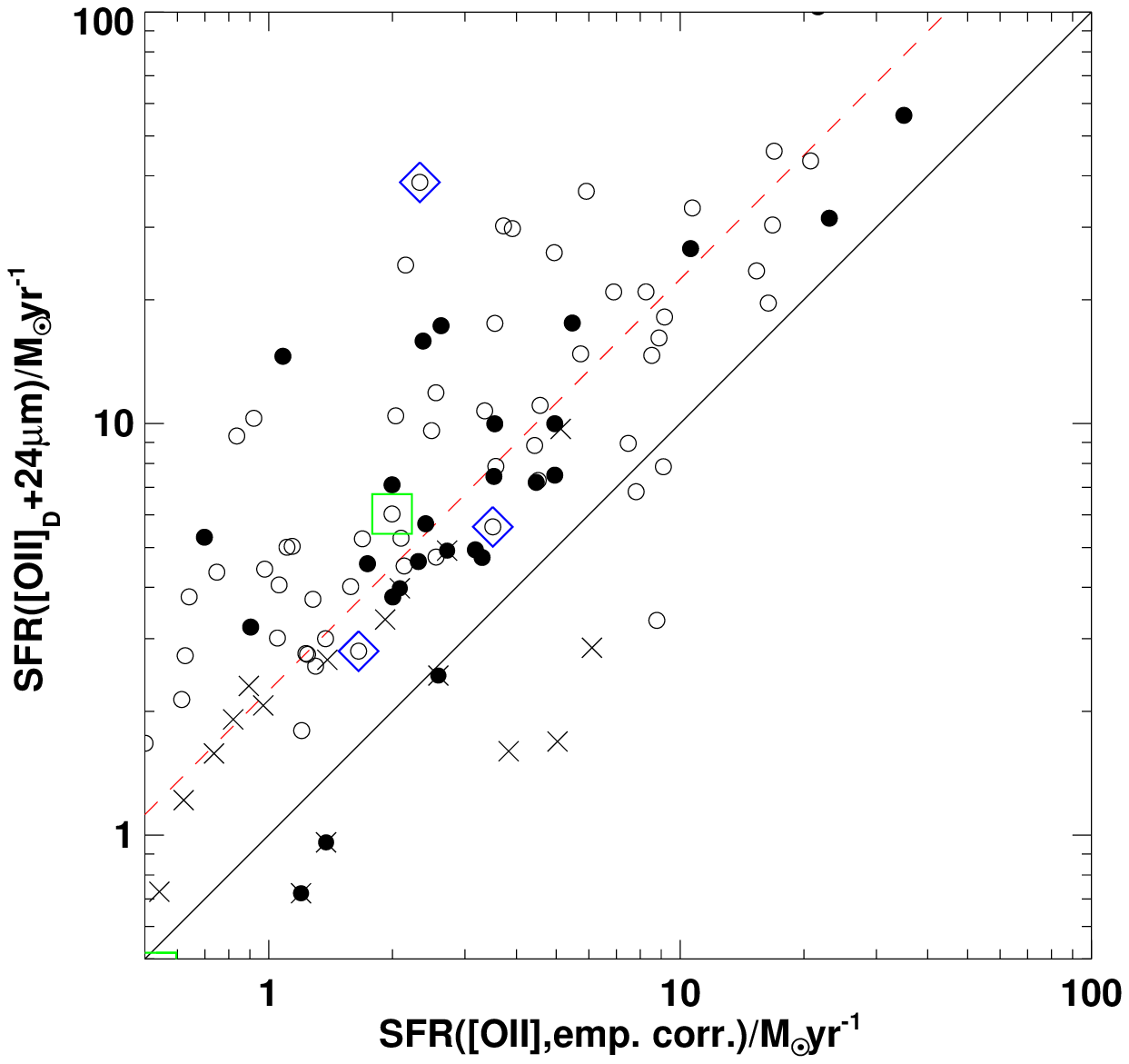}
	\caption{Comparison between empirically-corrected \oii-SFRs and \oii$+$24$\mu$m SFRs, as used by DEEP2/AEGIS.  Circles show all measurements with significant 24$\mu$m detections; crosses indicate S/N(24$\mu$m)$<$3. Filled circles show which galaxies are isolated by our automated criterion based on checking for [3.6]-detected neighbours in the $K$-selected FIREWORKS catalogue. Blue diamonds denote X-ray detections, which suggests that the \oii~and 24$\mu$m fluxes are contaminated by an AGN, and the green diamond indicates the possible presence of an AGN based on MIR colours (see Appendix \ref{sec:agn}). Note that the isolation criterion is likely too strict (as discussed in the text), and the majority of open circles are likely usable measurements, as supported by the fact that most open and filled circles follow the same average trend.}
	 \label{fig:cf24um}
}
\end{figure}

To understand any differences on our results from comparing our empirically-corrected \oii-SFRs with \oii$+$24$\mu$m SFRs from \citet{Noeske:2007tw}, the 24$\mu$m data are considered for the bright/high mass (and likely least-confused) FORS2 galaxies in CDFS, as described in \S\ref{sec:24data}. Keeping the previously mentioned caveats regarding source confusion in mind, Fig.~\ref{fig:cf24um} shows the comparison between \oii, $_{\rm emp. corr.}$-SFR and \oii$+$24$\mu$m SFRs, where the \oii~has not been empirically corrected, nor corrected for any extinction, before adding to the IR luminosity. For consistency with the DEEP2 method, the assumed \oii/\ha~ratio used here is 0.69 (rather than 0.5) and no correction for extinction is made (since the light from star-formation being reprocessed by dust should now be measured directly by the 24$\mu$m flux. The \oii~SFR using the DEEP2 assumptions is denoted SFR(\oii$_{\rm D}$) to be explicit. To convert the observed 24$\mu$m luminosity to total IR luminosity, the templates of \citet{Chary:2001fu}\footnote{using routines from: http://david.elbaz3.free.fr/astro\_codes/chary\_elbaz.html} are used (as used by \citealt{Noeske:2007tw}). The total IR luminosity is then converted to SFR using the prescription of \citet{Bell:2005qq} converted to our IMF. Filled circles show galaxies which are not considered confused, using the above (strict) criteria, and open circles show those which would be considered confused. Crosses indicate galaxies which do not have significant 24$\mu$m detections, and so the \oii$+$24$\mu$m SFR comes solely from \oii. As can be seen, the majority of these high mass objects show significant 24$\mu$m emission. The green square indicates a source which may be an AGN based on its MIR colours (see Appendix \ref{sec:agn}), so the 24$\mu$m flux may be contaminated by non-thermal emission. The blue diamonds denote objects which may contain an AGN based on X-ray emission (Appendix \ref{sec:agn}). The dashed line shows the best-fit relation, which is offset such that the SFR(\oii$_{\rm D}+24\mu m$) estimate is a factor of 2.2 higher than SFR(\oii, $_{\rm emp. corr.}$). Using instead the local conversion of  \citet{Rieke:2009gf} results in an even greater offset of the SFR including 24$\mu$m data. Without another independent estimate of the SFR, we cannot assess whether this offset is due to the empirical \oii~estimate underestimating the total SFR, or the \oii$+$24$\mu$m measurement overestimating it. To first order, we are primarily interested in the comparison between \oii-based SFRs at z$=$0.1 and z$=$1.0. The relative difference between the purely \oii-based measurements and those including 24$\mu$m data at z$=$1 will be considered below.

\subsubsection{Stacking analysis for low mass galaxies}
\label{sec:stack24}
Following the same method as for the higher mass FORS2 subsample above, isolated, lower-mass galaxies from ROLES are selected. Seven ROLES galaxies classed as isolated are individually detected in 24$\mu$m emission, and they follow the same relation, within the broad scatter, as the higher mass galaxies above. Considering only isolated galaxies without individual 24$\mu$m detections results in a sample of 32 objects. The 3$\sigma$ limit of the 24$\mu$m photometry in CDFS is 11$\mu$Jy \citep{Wuyts:2008lq}, which corresponds to a SFR of 0.7\msunyr~at z$=$1.  The fact that the majority of isolated ROLES galaxies are undetected down to this limit is reassuring that these low mass objects possess little dust, and hence the \oii~line should be tracing reliably their SFRs. In order to push to deeper limits, cut-outs are extracted from the 24$\mu$m image (obtained from the GOODS-S v0.30 release, Dickinson et al., in prep.) around the $K$-band position of the undetected objects and the mean of the stack is found. The total flux is measured in a series of apertures ranging from 3\arcsec~to 6\arcsec. No significant detection was made down to the limit of the image, which should be 2 (=11/$\sqrt{32}$) $\mu$Jy. Errors were calculated from 100 bootstrap resamplings of the 32 images used in the stack. The formal limit comes out to be $-0.17\pm0.24 \mu$Jy. This is many times lower than our \oii$_{\rm emp. corr}$ limit (SFR$>$0.3\msunyr), implying that the vast majority of star formation is unextinguished in these mass systems.

\begin{figure*}
	{\centering
		\includegraphics[width=85mm,angle=0]{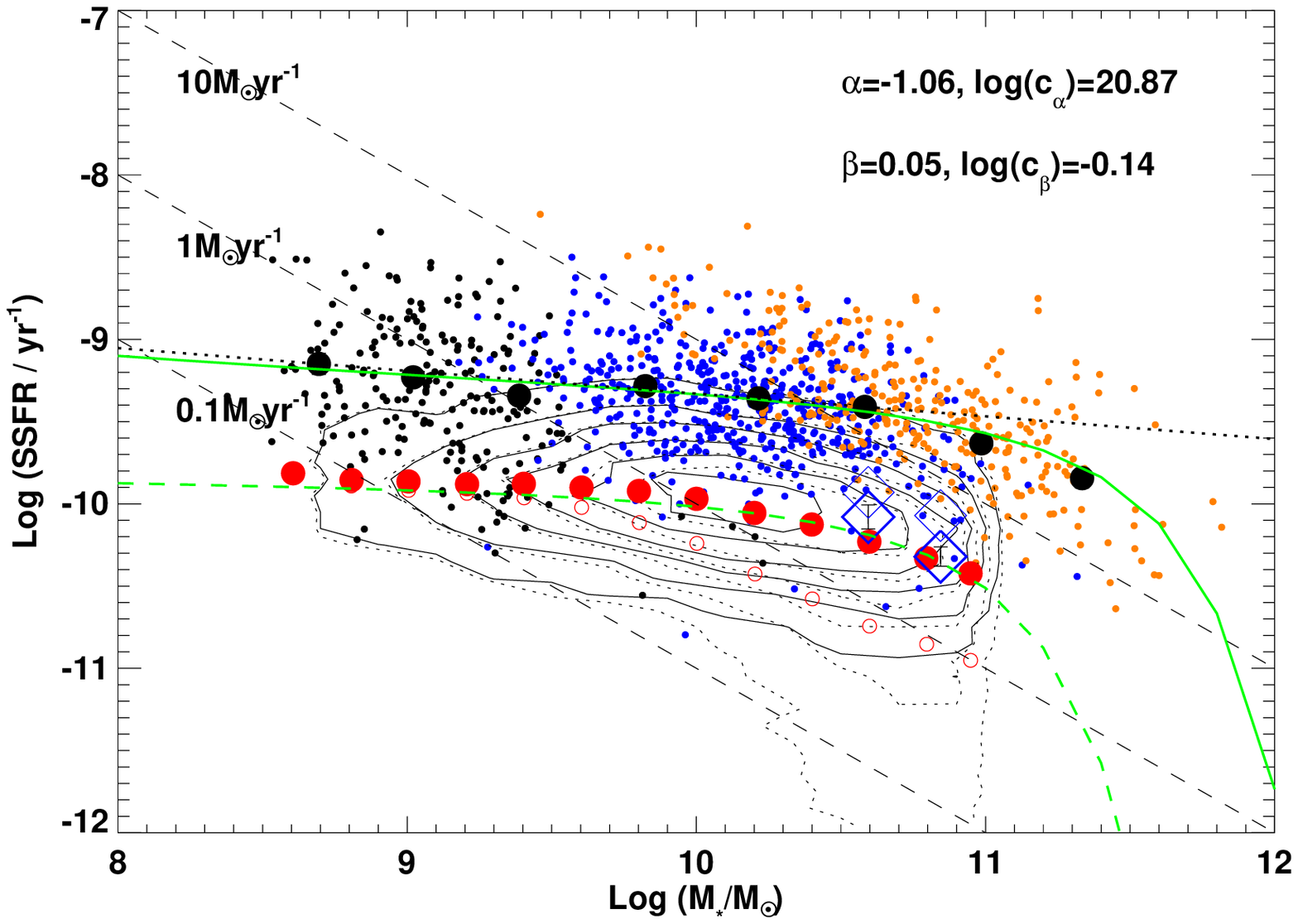}
		\includegraphics[width=85mm,angle=0]{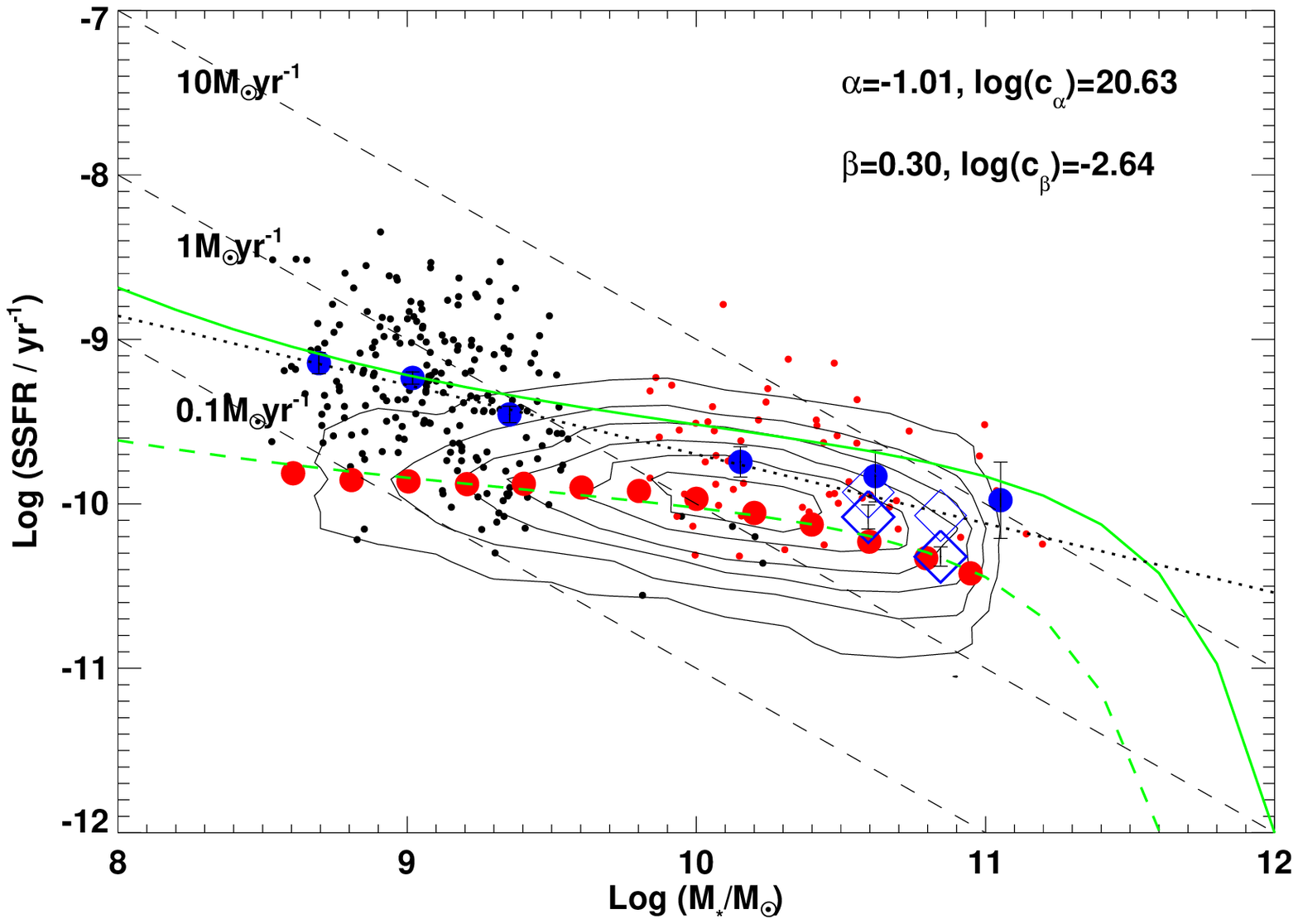}
	\caption{$\log$(SSFR) versus $\log$(stellar mass) for galaxies in our samples. Small, filled points show data at $0.88<z\le1.15$: black points are from ROLES, blue points are emission-line only SSFRs from DEEP2 and orange points are \oii$+$24$\mu$m SSFRs from DEEP2. Contours denote SDSS data at z$\sim$0.1. Larger filled circles show mean SSFRs in bins of stellar mass for the data in the two redshift ranges: black points at z$\sim$1 and red points at z$\sim$0.1. The dotted contours and open red circles show how the SDSS data would change if all galaxies with significant \ha~flux were included, rather than considering only blue cloud members, due to the long tail toward lower SSFRs at higher masses. Dashed lines denote the three different SFRs, as annotated. In the right panel, the higher mass data (smaller red filled points) are taken from ESO public spectroscopy rather than DEEP2. Thick blue diamonds with error bars show results taken from zCOSMOS ($0.7<z\le0.9$); thinner blue diamonds show these results evolutionary-corrected to $0.88<z\le1.15$ as described in the text. 
	Dotted black lines show best fit power laws to the z$=$1 data, as described in the text. Green lines show best-fit tau models, discussed in \S\ref{sec:tau}, with parameters as indicated on plot. See text for details.}
	 \label{fig:ssfrmass2}
}
\end{figure*}

\subsection{The SSFR--mass relation}
\label{sec:ssfrmass2}
As in Fig.~\ref{fig:ssfrmass}, SFRs and stellar masses from the z$\sim$0.1 and z$\sim$1 samples are combined to study the evolution of SSFR--mass in Fig.~\ref{fig:ssfrmass2}. The two panels show the different z$\sim$1 data. In both panels; the SDSS (z$\sim$0.1) galaxies are shown as contours for clarity. Filled red circles show the mean in bins of stellar mass. For the left panel, the effect of including all galaxies rather than just those which belong to the blue cloud is shown by the dotted contours with the mean relation as open red circles. This would give a lower average SSFR at higher masses, since higher mass galaxies possess a long tail toward lower SSFRs. The z$\sim$1 data are shown as smaller filled circles, colour coded by dataset: black points are ROLES' galaxies, blue points are blue, emission line only galaxies from DEEP2 and orange circles are DEEP2 \oii$+$24$\mu$m SFR measurements. Larger black circles show the median in bins of stellar mass for the z$\sim$1 data.  Dashed lines indicate SFRs of 0.1 \msunyr, 1 \msunyr, and 10 \msunyr. The black dotted line indicates a power law fit to the z$=$1 data which is given by $\log (SSFR) = -0.138 \log (M_\star) -7.95$. As can be seen, this fit is consistent with the bulk of the points but does not capture the turnover at the high-mass end, \ms$\gsim$10.5. [The higher mass points are better fit with  $\log (SSFR) = -0.577 \log (M_\star) -3.31$.]

The green curves show the best fit tau model (as described in \S\ref{sec:tau}) for these data, with parameters as annotated on the plot. The best fit using the DEEP2$+$ROLES data shows a much flatter dependence, almost zero, of $z_f$ on $M_b$ ($\beta=0.06$) than from the FORS2$+$ROLES data (\S\ref{sec:tau}, and repeated in right panel of Fig.~\ref{fig:ssfrmass2}).  The data points between 10$\lsim$\ms$\lsim$11 were not considered in \citet{Noeske:2007qa} due to their $<$95\% completeness but including or removing these makes little difference to the fit. 

The right hand panel of Fig.~\ref{fig:ssfrmass2} shows the same information, but the DEEP2 data have been replaced by the data from the public FORS2 spectroscopy in CDFS (smaller red filled circles). Recall that, for all these data, only blue cloud galaxies are used, and although individual data points are shown for each galaxy, the mean is calculated weighting each galaxy by a completeness and Vmax weight. For the FORS2 data, the mean relation is not significantly lowered by including all galaxies with \oii~emission, rather than just blue cloud galaxies with \oii~emission.

Clearly there is significant disagreement at the high mass end between the two different surveys used. To check that the difference between our ROLES$+$FORS2 dataset and the DEEP2/AEGIS data is not due to cosmic variance from our relatively small volume, we compare our measurements with the larger zCOSMOS survey. zCOSMOS SFRs from \oii~\citep{Maier:2009sf}, corrected to use our empirical mass-dependent correction, are taken. These data (plotted as thick open blue diamonds with error bars in the right panel of Fig.~\ref{fig:ssfrmass}) cover a slightly different redshift range ($0.7<z\le0.9$) from that of the ROLES and DEEP2 data. Estimating the size of this evolution by measuring the evolution from the DEEP2 data cut to the same redshift bins gives an increase of 0.15--0.25 dex between $0.7<z\le0.9$ and $0.88<z\le1.15$ in the two mass bins. This moves the open diamonds in Fig.~\ref{fig:ssfrmass} into closer agreement with the CDFS data (thinner blue diamonds). Thus, we are reassured that our high mass points in CDFS are not severely affected by cosmic variance.

At the high mass end, say \ms$\sim$10.6, the average SSFR measured in DEEP2 is 0.5 dex higher than that measured in CDFS or zCOSMOS. Part of this discrepancy may be due to the use of different SFR indicators. Fig.~\ref{fig:cf24um} shows that using SFR(\oii$_{\rm D}$+24$\mu$m) instead of SFR(\oii),$_{\rm emp. corr.}$ leads to a factor of 2.2 (0.35dex) higher estimate of the SFR. \citet{Kelson:2010ul} have suggested that using observed 24$\mu$m at z$\sim$1 to estimate SFRs may overestimate the ongoing SFR by factors of $\sim$1.5-6 since dust-encircled thermally pulsating-asymptotic giant branch (TP-AGB) stars with ages of 0.2-1.5Gyr may constitute a significant source of contamination to flux interpreted as coming from star-formation. Early results from the {\it Herschel Space Observatory} show that there is considerable scatter between TIR luminosities measured directly from near the IR peak of the SED (at 100$\mu$m and 160$\mu$m) and those extrapolated from observed 24$\mu$m luminosities, with a systematic bias which is likely a function of redshift \citep{Rodighiero:2010ve}. \citet{Nordon:2010fk} find that {\it Spitzer} 24$\mu$m observed SFRs overestimate {\it Herschel}-measured SFRs by factors of $\sim$4-7.5 at slightly higher redshift (1.5$<$z$<$2.5). The details of these likely depend on the SED-fitting method to both datasets, and it is not trivial to see how easily these results transfer to the data considered here. However, even considering just the DEEP2 points using only \oii~SFRs (generally lower mass, \ms$\sim$10) shows a significantly higher SFR in the\oii~data versus the CDFS data, and so the difference cannot be attributed entirely to the inclusion of the 24$\mu$m data.

One main difference between the DEEP2 \oii~data and the other two datasets is that DEEP2 is not flux calibrated (as the other samples are), but line strengths are calculated using equivalent widths and broad band photometry to calibrate the continuum. Another difference is that DEEP2 is very biased towards blue-selection relative to the other surveys (CDFS is $z$-band selected, zCOSMOS is $I$-band selected). Although the $K$-band follow-up photometry helps with the mass selection of the AEGIS-DEEP2 data, the initial DEEP2 spectroscopy is still $R$-band selected (which is close to rest-frame $B$ at z$\sim$1, and thus more closely resembles a SFR-seleted sample rather than mass-selected). Thus, galaxies with low SFRs might possibly be missed, leading to an absence of low SSFR objects. In practice, it may be that a subtle interplay of factors are responsible for the disagreement of the DEEP2 data with the other results. Indeed, \citet{Cowie:2008ob} make a similar suggestion when arguing why they find a much longer tail toward lower SSFRs compared with the DEEP2 results.

\section{AGN}
\label{sec:agn}

\begin{figure}
	{\centering
	\includegraphics[width=85mm,angle=0]{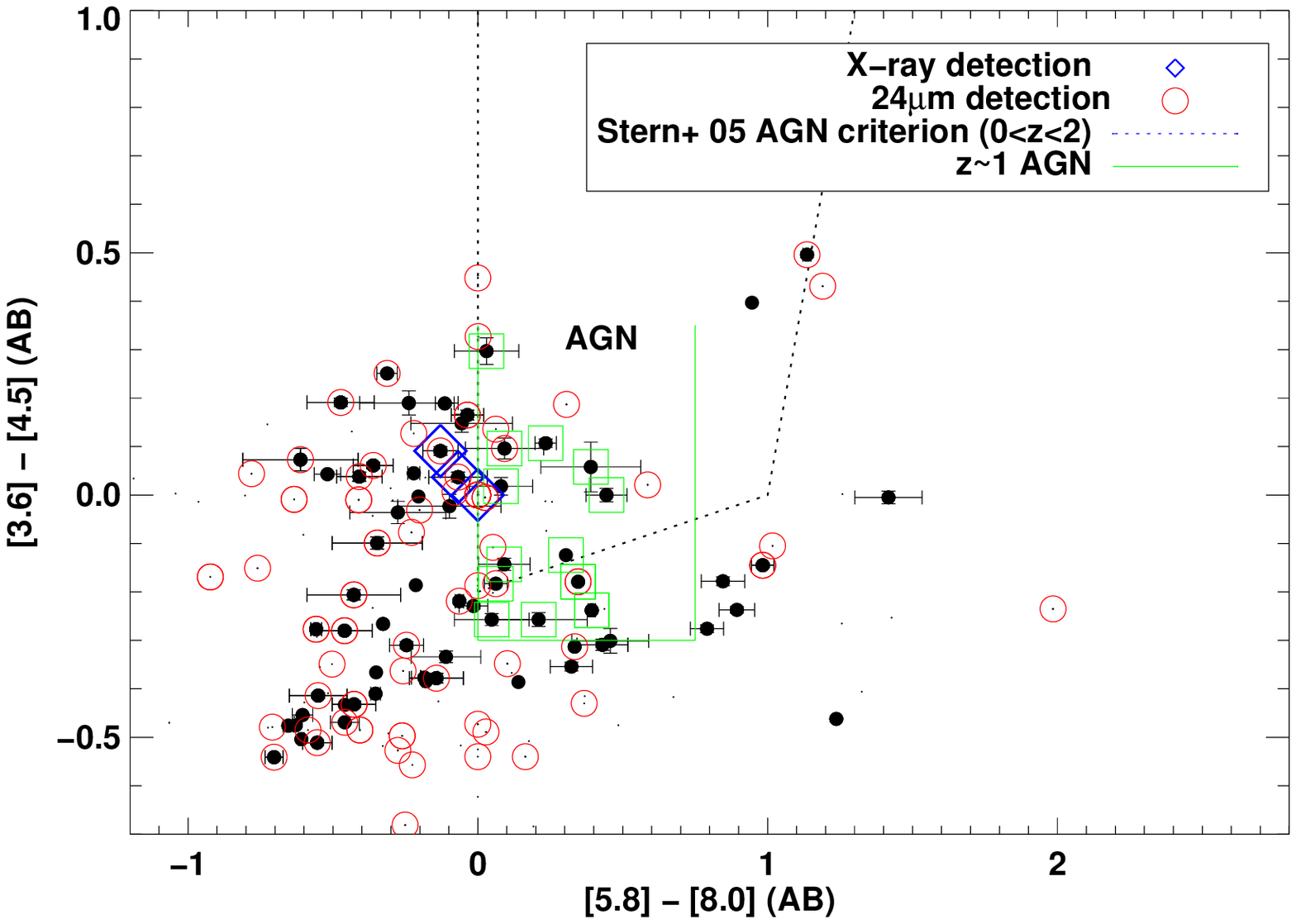}
	\caption{MIR colour--colour diagram to examine for the presence of AGN. Points are taken from both ROLES and FORS2 spectroscopy in CDFS. Objects with $>5\sigma$ detections in all four filters are shown as filled circles with error bars. Red open circles indicate $>3\sigma$ 24 $\mu$m detections. The dotted line indicates the region populated by AGN according to the general $0<z<2$ criterion of \citet{Stern:2005eu}. The solid green box shows a refinement of this designed to target our redshift range. Blue diamonds indicate X-ray detected galaxies.}
	 \label{fig:agn}
}
\end{figure}

In order to test for contamination of the \oii~flux or 24$\mu$m flux by AGN in our sample, the technique of \citet{Stern:2005eu} is used to examine the MIR colours for the signature of an AGN. [5.8]-[8.0] vs [3.6]-[4.5] colours for all galaxies in both the ROLES and FORS2 subsamples of CDFS are shown in Fig.~\ref{fig:agn}. Objects with greater than 5$\sigma$ detections in each of the four bands are indicated as filled circles with error bars. Only these significant detections are considered when testing for the presence of an AGN, as reliable colours cannot be measured for the other objects. The dotted line indicates the region within which AGN should lie according to the criteria of \citet{Stern:2005eu} for galaxies over the redshift range $0<z<2$. We refine these criteria slightly considering the boundary around AGN-like objects in the redshift range $0.8\lsim z \lsim 1.2$ by reading values off figs.~2 \& 3 of \citet{Stern:2005eu} which bracket the majority of the AGN population. This results in the green lines plotted in Fig.~\ref{fig:agn}. As can be seen, only a small minority of galaxies (with small photometric errors) lie within this region . AGN following these criteria are indicated by green squares. Only 15 galaxies out of our total sample of 311 meet this definition, and only two of these are drawn from the FORS2 sample, plotted in Fig.~\ref{fig:cf24um}. It is also instructive to examine which of these objects are X-ray luminous and thus classified as AGN this way. Four such X-ray detections are present in the combined ROLES$+$FORS2 sample (three from FORS2; one from ROLES). Of these four, one is undetected in all four IRAC bands (indicated as the point at 0,0 in the plot), and one is only significantly detected at [5.8]. The remaining two are shown in the plot and lie just outside our z$\sim$1 AGN criteria: one is within 1$\sigma$ of the edge, and the other just over 2$\sigma$ away. This suggests that the MIR colour is not an unreasonable method of selecting AGN. More importantly, this test suggests that contamination of our SFR estimates by AGN is likely to be minimal, and we are justified in neglecting it in our analysis.

\label{lastpage}

\end{document}